\documentclass[a4paper]{article}

\usepackage{mathtools}
\usepackage{ wasysym }
\usepackage[dvipsnames]{xcolor}
\usepackage{color}
\usepackage{tikz}
\usetikzlibrary{shapes}
\usepackage{lineno}
\usepackage{babel}
\usepackage{hyperref}
\usepackage{eufrak}
\usepackage{mathrsfs}
\usepackage[linesnumbered, ruled, vlined]{algorithm2e}
\usepackage{amsfonts}

\definecolor{mygrey}{rgb}{0.3, 0.3, 0.3}
\definecolor{mylightgrey}{rgb}{0.7, 0.7, 0.7}

\newcommand{\inner}[2]{\left< #1, #2 \ \right>}

\begin{document}

%%%% Article title to be placed here
\title{Impact of a rigid sphere onto an elastic membrane}

\author{%%%% Author details
Elvis A. Agüero$^{1}$, Luke Alventosa$^{2}$, Daniel M. Harris$^{2}$ \\ and Carlos A. Galeano-Rios$^{3,*}$\\ \\
%%%%%%%%% Insert author address here
$^{1}$ILACVN, Universidade Federal da Integra\c{c}\~{a}o Latino-Americana,\\ Foz de Igua\c{c}u, 85867-970 PR, Brazil.\\
$^{2}$School of Engineering, Brown University, Providence,\\ RI 02912, USA.\\
$^{3}$School of Mathematics, University of Birmingham, \\ Birmingham B15 2TT, United Kingdom.\\ \\
*grcarlosa@gmail.com}
\date{}
\maketitle
%%%% Abstract text to be placed here %%%%%%%%%%%%
\begin{abstract}
We study the axisymmetric impact of a rigid sphere onto an elastic membrane theoretically and experimentally. We derive governing equations from first principles and impose natural kinematic and geometric constraints for the coupled motion of the sphere and the membrane during contact. The free-boundary problem of finding the contact surface, over which forces caused by the collision act, is solved by an iterative method. This results in a model that produces detailed predictions of the trajectory of the sphere, the deflection of the membrane, and the pressure distribution during contact. Our model predictions are validated against our direct experimental measurements. Moreover, we identify new phenomena regarding the behaviour of the coefficient of restitution for low impact velocities, the possibility of multiple contacts during a single rebound, and energy recovery on subsequent bounces. Insight obtained from this model problem in contact mechanics can inform ongoing efforts towards the development of predictive models for contact problems that arise naturally in multiple engineering applications.\\ \\
{Keywords: Impacts, waves, contact, membrane, rebound, elasticity}
\end{abstract}

%%%%%%%%%%%%%%%%%%%%%%%%%%%

%%%%%%%%%% Insert the texts which can accomdate on firstpage in the tag "fmtext" %%%%%

%%%%%%%%%%%%%%% End of first page %%%%%%%%%%%%%%%%%%%%%

\section{Introduction}
%%%% Insert A head here
Mechanical contact problems arise naturally in countless industrial and scientific applications. Classical examples include the study of the deformation and stress in gear teeth \cite{bruzzone20212d,bruzzone2021gear,rosso2019proposal}, ball bearings and ball joints \cite{askari2021mathematical}, impact absorbers \cite{hundal1976impact},  propagation of stress waves in colliding solids \cite{Johnson1987}, and models of granular materials \cite{cundall1979discrete}.  Contact problems also frequently arise in problems relating to material characterisation, where a localised indenter is used to infer properties of solid substrates \cite{herbert2001measurement}. In particular, the mechanical response of elastic films and membranes under indentation \cite{begley2004spherical,komaragiri2005mechanical,selvadurai2006deflections} has seen recent interest for applications in characterising soft polymeric and biological materials \cite{ahearne2005characterizing}, or `2D materials' such as graphene \cite{gupta2015recent}.

Several studies in contact mechanics have stemmed from the seminal works of Hertz \cite{Hertz1896,Hertz1882beruhrung}; in which, friction effects are neglected, the shape of the contact region (in the vicinity of the initial contact point) is approximated by a paraboloid, and the resulting contact surfaces are elliptical. The work of Hertz covered mainly static contacts, yielding predictions that have held remarkably well. Moreover, Hertz also considered impacts of deformable solids, working within the framework of quasi-static approximations. In particular, waves generated by the impact were ignored \cite{Hertz1896,Johnson1982}.

\subsection{Non-Hertzian problems}
Problems that do not conform to the simplifying hypotheses of Hertz are called non-Hertzian. The solution to contact problems of this kind involves a free-boundary problem on 2-dimensional (2D) surface, i.e. finding the bounding curve of the portion of the outer surface of the solids, where contact happens. Moreover, this free-boundary problem is embedded within a 3D free-boundary problem, i.e. finding where the deformed external surfaces of the contacting solids lie in the first place.

These nested free-boundary problems are also coupled; as the extent of the pressed surface influences the pressure distribution, which in turn influences the shape of the solids, on whose outer surface lies the contact surface. The coupling of these free-boundary problems brings in non-linearities of geometric origin, even when the partial differential equations that govern the deformation of the solids are linear.

Non-Hertzian contact problems represent a substantial number of cases of interest in engineering applications. Due to their complexity, analytical solutions are often unavailable and, thus, they are typically tackled using numerical methods\cite{Karami1989non}. The nested free-boundary problems they involve are solved using strategies that include imposing energy minimisation principles \cite{KalkerAndRanden1972}, and iterating on the extent of the pressed surface until the pressures obtained are all positive and there is no superposition of the solids outside the region where the pressure is applied \cite{PaulAndHashemi1981}.

\subsection{The kinematic match}
The model problem of the impact of a rigid sphere onto a deformable substrate has proven to be very useful in the study the dynamic behaviour of deformable bodies that undergo a collision \cite{LeeAndKim2008, CourbinEtAl2006,EichwaldEtAl2010,Galeano-RiosEtAl2017,Galeano-RiosEtAl2019,Galeano-RiosEtAl2021}. Moreover, the transfer of energy to waves during contact was shown to be successfully captured with relatively simple models in the same set-up \cite{Galeano-RiosEtAl2021}. Furthermore, in \cite{Galeano-RiosEtAl2017,Galeano-RiosEtAl2021}, the use of moving meshes or variational methods, such as the finite element method, was not strictly necessary to solve these type of impacts (though there is, in principle, no impediment to use them); instead, it was sufficient to use the finite difference method, which is easier to program and, therefore, accessible to a larger community of modellers.

The kinematic match (KM) method, introduced in \cite{Galeano-RiosEtAl2017} was first developed as a fully-predictive method to solve the impact and rebound of a rigid hydrophobic sphere onto the free surface of a bath, an application for which it has been successfully validated using experimental data (see figures 3a and 11 in \cite{Galeano-RiosEtAl2017}; figures 5, 7 and 9 in \cite{Galeano-RiosEtAl2019}; and figures 6a and 7 in \cite{Galeano-RiosEtAl2021}), as well as direct numerical simulations (see figures 6b, 7, 9 and 12 in \cite{Galeano-RiosEtAl2021}). However, the method has far more general applications. In particular, the matching conditions imposed by the KM are agnostic in relation to the type of equations that govern the motion of the impacting surfaces. This fact strongly suggests that the modelling of simplified impact problems with the KM can inform future work in diverse engineering applications of significant complexity.

The KM imposes only the most natural kinematic and geometric constraints to the motion of the impacting surfaces. Some of the conditions it imposes are common to those already used in contact mechanics of solids; however, unlike other contact-mechanics methods, the KM introduces a tangency condition at the boundary of the contact surface. The matching conditions imposed yield the equations needed to solve the free-boundary problem of finding the pressed surface and the pressure distribution supported on it. Motivated by the virtues of the KM, we here present a first effort to apply the KM framework to the solution of non-Hertzian contact problems of solids. The method promises to be of importance for problems where dynamic effects are relevant; in particular, those for which the effects of waves caused by collisions are non-negligible.

\subsection{The model problem}
In the present work, we consider the problem of a rigid sphere impacting on an elastic membrane and we formulate its mathematical representation along the lines of the KM method. Moreover, we improve the original form of the KM, expanding the compatibility conditions in the pressed area, while also reducing the size of the resulting system of equations. The resulting equations are solved numerically, yielding predictions for the contact time, trajectory of the impactor, deflection of the membrane, coefficient of restitution of the impacting sphere, and the evolution of the pressed surface as well as the pressure distribution supported on it.

In some cases, it is possible that a decidedly simpler quasi-static model may be appropriate to model impacts on an elastic membrane.  We anticipate this to occur when the kinetic energy of the membrane during the impact process is negligible as compared to its elastic energy.  Such kinetic and elastic energies can be estimated to leading order by $E_k \sim \mu \Lambda^2 (\delta/t_c)^2$ and $E_e \sim \tau \delta^2$ \cite{CourbinEtAl2006}, respectively, where $\mu$ is the area density of the membrane, $\Lambda$ is the membrane radius, $\tau$ is the membrane tension, and $\delta$ is the maximum deflection of the membrane during an impact occurring over a time scale $t_c$.  By requiring $E_k \ll E_e$ we find the condition 
\begin{equation}\label{eqn:qs_scale}
    \frac{\tau t_c^2}{\mu \Lambda^2} \gg 1
\end{equation}
that thus corresponds to the quasi-static limit.  

In a tensioned membrane the wave speed is known to be $C=\sqrt{\tau/\mu}$, and thus a timescale for wave propagation in the membrane can be defined as $t_p=\Lambda/C$.  Upon substitution, our quasi-static condition (\ref{eqn:qs_scale}) can also be reinterpreted as a ratio of time scales:
\begin{equation}\label{eqn:qs_time_scales}
    \frac{t_c^2}{t_p^2} \gg 1,
\end{equation}
or that the timescale of impact must be sufficiently long as compared to the time scale of wave propagation.  Finally, for a freely impacting mass $m$, the contact time on a membrane of constant tension scales like $t_c \sim \sqrt{m/\tau}$ \cite{GiletAndBush2009}, and thus our condition can also be rewritten as a mass ratio: 
\begin{equation}\label{eqn:massratio}
    \frac{m}{\mu \Lambda^2} \gg 1,
\end{equation}
or that the mass of the impactor is much greater than the total mass of the membrane.
Should the our impact parameters occur outside of this limit, we expect dynamic processes to be important in determining the subsequent dynamics, requiring a non-Hertzian model.

An experimental set-up was designed to test the predictions produced by our model against controlled experiments. Our predictions for contact time and maximum surface deflection match our experimental results remarkably well, while also being in line with prior experimental results reported in \cite{CourbinEtAl2006}.

Section \ref{section:Problem_formulation} presents the rigorous mathematical formulation of the impact problem, including the matching conditions of the KM method. Section \ref{section:Numerical_implementation} details the numerical approximations and schemes, as well as the iterative method used to capture the moving boundary of the contact area. The experimental set-up and procedures are detailed in section \ref{section:Experiments}. Comparisons to our experimental data, together with other predictions of the model here introduced are presented in section \ref{section:Results}. We discuss the implications of our findings and describe ongoing directions of development in section \ref{section:Discussion}. Julia, Python and Matlab codes, used for the computational implementations of the methods here presented are made available in a public repository, while videos of the experiments, and animations of the results are made available as supplementary material.

\section{Problem formulation}\label{section:Problem_formulation}
\begin{figure}
    \centering
    \includegraphics[width=.5\textwidth]{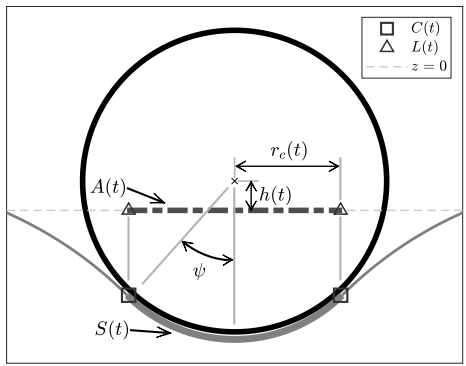}
    \caption{Schematic representation of the impact. The elastic membrane is shown with thin grey solid lines outside the contact surface $S(t)$, and with a thick grey solid line inside $S(t)$. It should be noted that, in this model $S(t)$ is a subset of the graph of $\eta(t)$, and that the separation shown is merely for illustrative purposes. The orthogonal projection of $S(t)$ onto the $(r,\theta)$-plane, $A(t)$, is shown with a thick dark grey dashed line. Curves $C(t)$ and $L(t)$, which respectively bound $S(t)$ and $A(t)$, are seen as points in this cross-section. Variables $h(t)$ and $r_c(t)$ correspond to the height of the centre of mass of the sphere and the radius of $A(t)$, respectively.}
    \label{fig:Scheme}
\end{figure}

We consider the case of an elastic membrane of mass per unit area $\mu$, supported by a circular rim of radius $\Lambda$, and subject to initial isotropic stress $\tau$ (see figure \ref{fig:Scheme}). We introduce cylindrical coordinates $(r,\theta,z)$ with the rim on the $z = 0$ plane, the origin at the centre of the rim, and gravity given by $\vec{g} = -g\hat{z}$. 

At time $t=0$, this elastic membrane lies in equilibrium, deformed by the action of its own weight, as the lowest point (the "south pole") of a homogeneous rigid sphere of radius $R$ and mass $m$, that moves with a velocity $\Vec{v}(t = 0) = -V_0\hat{z}$, is in imminent contact with the centre of the mesh (i.e. the height of the south pole coincides exactly with the height of the centre of the at-rest membrane).

We will consider only axisymmetric impacts in the present work and, therefore, we ignore all dependence on the $\theta$ variable from here on. Non-axisymmetric impacts can also be modelled by the methods here introduced; however, these will be the subject of a separate article.

\subsection{Governing equations}
We take $R$, $C = \sqrt{\tau / \mu}$ and $P= \tau / R$ as the characteristic length, velocity and pressure, respectively; and we define the following dimensionless numbers
\begin{equation}
\label{eqn:dimensionless_numbers}
    \mathfrak{F} \coloneqq \frac{g\mu R}{ \tau}, \qquad \mathfrak{L} \coloneqq \frac{\Lambda}{R}, \qquad
    \mathfrak{U} \coloneqq \frac{V_0}{C}, \qquad 
    \mathfrak{M} \coloneqq \frac{\mu R^2}{m}.
\end{equation}
We make the simplifying assumptions that the displacement of all points in the membrane happens exclusively along the $z$ direction and that the tension on the membrane $\tau(r,t)$ is constant everywhere and throughout the impact.

We define $\eta(r, t) : [0, \mathfrak{L}] \times [0, +\infty) \to \mathbb{R}$ as the deflection of the membrane, and we introduce the vertical surface velocity $u(r,t) \coloneqq \partial_t\eta$.  Disregarding friction between the sphere and the membrane, the effect of the impact can be modelled by a pressure distribution $p = p(r,t)$ supported on $A(t)$, the orthogonal projection of the contact surface $S(t)$ onto the $(r,\theta)$-plane (see figure \ref{fig:Scheme}). 

We define $\kappa = \kappa(\eta)$ as twice the mean curvature of membrane, i.e.
\begin{equation}
    \kappa 
    = 
    \frac{
    \partial_{rr}\eta
    }{
    \left[
    1
    +
    \left(\partial_{r}\eta\right)^2
    \right]^{\frac{3}{2}}
    }
    +
    \frac{
    \partial_{r}\eta
    }{
    r
    \left[
    1
    +
    \left(\partial_{r}\eta\right)^2
    \right]^{\frac{1}{2}}
    };
\end{equation}
and thus the elevation of the membrane is governed by
\begin{eqnarray}
\label{eqn:wave_non_lin}
    \partial_{tt}
    \eta 
    =
    -
    \mathfrak{F} +
    \kappa
    -
    p, 
    &&
    \forall\ (r,t)\in [0,\mathfrak{L}]\times(0,\infty),
\end{eqnarray}
subject to
\begin{eqnarray}
    p(r,t) = 0,
    &&
    \forall \ 
    r,t;r\notin A(t),
    \\
    p(r,t=0)=0,
    &&
    \forall \ r \in [0,\mathfrak{L}],
    \\
    \label{eqn:mbn_boundary_cnd_1}
    \eta(\mathfrak{L}, t) 
    =
    0,
    &&
    \forall\ 
    t
    \geq
    0,
\end{eqnarray}
with the initial conditions given by
\begin{eqnarray}
    \label{eqn:initial_cond_eta}
    \kappa(\eta(r,t = 0))
    =
    \mathfrak{F},
    &&
    \forall r \in [0,\mathfrak{L}]
\end{eqnarray}
and 
\begin{eqnarray}
    u(r,t=0) 
    =
    0, 
    &&
    \forall\ 
    r
    \in
    [0,\mathfrak{L}].
\end{eqnarray}

We note that equation (\ref{eqn:initial_cond_eta}) imposes that, as the membrane is about to be hit, it is found at its equilibrium shape, as dictated by its own weight distribution and initial tension. See Appendix \ref{App:Wave_eq} for a summarised derivation of equation (\ref{eqn:wave_non_lin}), which also applies to (\ref{eqn:initial_cond_eta}), as the steady state form of (\ref{eqn:wave_non_lin}).

We define $h = h(t)$ as the $z$ coordinate of the centre of mass of the sphere and $v(t) \coloneqq h'(t)$. By Newton's second law we have
\begin{eqnarray}
\label{eqn:sphere_non_dim}
    v'(t)
    =
    -
    \mathfrak{F}
    +
    \mathfrak{M}
    \int\limits_{A(t)}
    {p\, dA},
    &&
    \forall \ t\in(0,\infty)
\end{eqnarray}
subject to
\begin{equation}
    v(t=0) = - \mathfrak{U}
\end{equation}
and
\begin{equation}
    h(t=0) = 1+\eta(r=0,t=0).
\end{equation}

\subsection{The kinematic match}
%%%% Insert B head here
Four compatibility conditions are imposed. First, on the contact area $A(t)$, the two surfaces must coincide, that is
\begin{eqnarray}
    \eta(r,t) = h(t)+s(r)
    &&
    \forall
    r,t; r\in A(t),
\end{eqnarray}
where $s$ is given by the expression for lower hemisphere of the unit sphere centred at the origin, that is
\begin{equation}
    s(r)
    =
    -
    \left|
    \sqrt{(1-r^2)}
    \right|, 
    \quad
    \forall \ 
    r\in [0,1].
\end{equation}
Secondly, the velocity of the membrane $u$ must satisfy
\begin{eqnarray}
\label{eqn:velocity_match}
    u(r,t) = v(t), 
    &&
    \forall\ 
    r,t; r\in A(t).
\end{eqnarray} 

Here, we are implicitly assuming that the deformation of the elastic membrane is such that the surface can be described at all times by a well-defined function of $r$. This assumption is extremely reasonable for the present case, though it is not strictly required by the KM formulation. Were it violated, the conditions above could be formulated in terms of the contact surface $S(t)$ to allow the method to be applicable. Furthermore, we assume here that the contact surface changes continuously.

The third condition requires that, at the boundary of the contact surface $S(t)$ (i.e. the contact curve $C(t)$), the deformable surface be differentiable. This is equivalent to
\begin{eqnarray}
\label{eqn:velocity_constraint}
    \partial_{r}
    \eta(r,t)
    =
    s'(r),
    &&
    \forall\ 
    r,t; r \in L(t),
\end{eqnarray}
where $L(t)$ is the orthogonal projection of $C(t)$ onto the $(r,\theta)$-plane (see figure \ref{fig:Scheme}).

Our final compatibility condition requires that there be no superposition between the sphere and the membrane outside of the pressed surface, which is equivalent to
\begin{eqnarray}
    \eta(r,t)
    <
    h(t)
    +
    s(r),
    &&
    \forall\ 
    r\leq 1, r \notin A(t).
\end{eqnarray} 

We note that, in prior formulations of the kinematic match \cite{Galeano-RiosEtAl2017,Galeano-RiosEtAl2019,Galeano-RiosEtAl2021}, the condition given by equation (\ref{eqn:velocity_match}) was not included. Instead, the implementation relied on including equations that satisfied this condition approximately. The current choice is both, more efficient and more accurate, besides being more physically intuitive.

\subsection{Further simplifications}
In the present problem, we further assume that the pressed area is simply connected; which, within our axially symmetric configuration, is equivalent to $S(t)$ being a spherical cap centred at the lowest point of the sphere and, consequently, $A(t)$ being a circle of radius $r_c(t)$.

Moreover, we note that, in the equations above, the only non-linear term is the one given by $\kappa=\kappa(\eta)$, in equation (\ref{eqn:wave_non_lin}); however, this is by no means the only non-linearity in the problem. The other source of non-linearity is "hidden" in the problem of finding the pressed area. 

As the main focus of the present work is to present a first application to solid mechanics of a method to solve the non-linearity that is intrinsically embedded in this type of impact problems, we direct the focus of our presentation to this specific non-linearity; the solution of which, as will be shown in section \ref{section:Numerical_implementation}, can be achieved by an iteration on the geometry of the contact surface. Consequently, we choose to linearise the curvature function where needed. Otherwise, we would be forced to used nested iterations, unnecessarily obscuring the presentation of the main ideas here considered.

We note that, inside the pressed surface (i.e. $\forall \ r,t; r \leq r_c(t)$), we do not need to linearise the curvature operator $\kappa$, as we know that in this region $\kappa = 2$, since it is given by two times the reciprocal of the dimensionless radius of the sphere.

We thus define
\begin{equation}\label{eqn:kappa}
    \check{\kappa}(\eta;r_c)
    \coloneqq
    \left\{
    \begin{array}{cc}
        2 &  r \leq r_c(t),\\
        \partial_{rr} \eta + \frac{1}{r} \partial_{r} \eta & r > r_c(t).
    \end{array}\right.
\end{equation}

\subsection{Summarised model}
The axisymmetric impact of a solid sphere at the centre of a tensioned circular membrane is thus modelled by the solution to
\begin{eqnarray}
\label{eqn:wave_non_lin_01}
    \label{eqn:eta_t}
    \partial_t \eta = u,
    &&
    \forall\ (r,t)\in [0,\mathfrak{L}]\times(0,\infty),
    \\
    \label{eqn:membrane}
    \partial_{t}
    u
    =
    -\mathfrak{F} +\check{\kappa}-p, 
    &&
    \forall\ (r,t)\in [0,\mathfrak{L}]\times(0,\infty),
    \\
    h'(t) = v(t), 
    &&
    \forall \ t\geq 0,
    \\
    \label{eqn:sphere_non_dim_01}
    v'(t)
    =
    -
    \mathfrak{F}
    +
    \mathfrak{M}
    \int\limits_{A(t)}
    {p\, dA},
    &&
    \forall \ t\geq 0,
\end{eqnarray}
subject to
\begin{eqnarray}
    \label{eqn:mbn_boundary_cnd}
    \eta(\mathfrak{L}, t) = 0,
    &&
    \forall\ t \geq 0,
    \\
    \partial_{r}\eta(r=0,t) = 0,
    &&
    \forall\ t \geq 0,
    \\
    \label{eqn:initial_cond_eta_01}
    \check{\kappa}(\eta(r,t = 0)) = \mathfrak{F},
    &&
    \forall r \in [0,\mathfrak{L}],
    \\
    u(r,t=0) = 0, 
    &&
    \forall\ 
    r
    \in
    [0,\mathfrak{L}],% \\ \partial_{r}u(r=0,t) = 0, &&\forall\ t\geq 0,
    \\
    p(r,t) = 0,
    &&
    \forall \ r,t;r>r_c(t),
    \\
    p(r,t=0) = 0,
    &&
    \forall\ r\in[0,\mathfrak{L}],
    \\
    h(t=0) = 1+\eta(r=0,t=0),
    &&
    \\
    v(t=0) = - \mathfrak{U},
    &&
    \\
    \eta(r,t) = h(t)+s(r),
    &&
    \forall
    r,t; r\leq r_c(t),
    \\
    u(r,t) = v(t), 
    &&
    \forall\ 
    r,t; r\leq r_c(t),
    \label{eqn:velocity_constraint_01}
    \\
    \label{eqn:tangency_constraint}
    \partial_{r}
    \eta(r_c(t),t)
    =
    s'(r_c(t)),
    \\
    \label{eqn:no_overlap}
    \eta(r,t)
    <
    h(t) + s(r),
    &&
    \forall\ r;
    r_c(t)<r\leq 1;
\end{eqnarray}
where we are using the symmetry of the problem at the origin to define boundary conditions.

We note that, once the dimensionless numbers $\mathfrak{F}$, $\mathfrak{L}$, $\mathfrak{M}$ and $\mathfrak{U}$ are given, the impact problem is completely defined.

\subsection{Quasi-static model}\label{sec:qs}

In the quasi-static limit discussed earlier, the free membrane (outside of the contact region) satisfies Laplace's equation $\nabla^2 \eta = 0$
to linear order for $r_c \leq r \leq \mathfrak{L}$ (neglecting the weight of the elastic sheet). Recall that $r_c= \sin \psi$ is the radius of contact between the sphere and membrane (see figure \ref{fig:Scheme}), $\eta$ is the deflection of the membrane, and all lengths non-dimensionalised by $R$.  Under these assumptions, the deformation has a known analytical solution (with the outer boundary of the membrane fixed such that $\eta(\mathfrak{L}) = 0$),
  \begin{equation}
     \eta (r) = A_0 \ln \left ( \frac{r}{\mathfrak{L}} \right).
 \end{equation}
 To determine $A_0$, the tangency boundary condition at the point of contact is applied. In other words,
 \begin{align}
     \partial_r \eta(r_c) &= \tan \psi. 
 \end{align}
 Thus the solution for the membrane shape becomes 
 \begin{equation}
     \eta (r) = r_c \tan \psi \ln \frac{r}{\mathfrak{L}}.
 \end{equation}
 Now, we need to determine the radius of contact, $r_c$, that occurs when a sphere is resting statically on the membrane and displaces the center of the membrane by an amount $\delta_s$.  We can thus write
 \begin{equation}
     \delta_s = -\eta(r_c) + (1-\cos \psi) = -r_c \tan \psi \ln \frac{r_c}{\mathfrak{L}} + (1-\cos \psi).
     \label{delta}
 \end{equation}
 This algebraic equation can be solved numerically for $r_c$ for each $\delta_s$.

Furthermore, in this limit, equations (\ref{eqn:kappa}) and (\ref{eqn:membrane}) imply that $p=2$ in the contact region, and thus the trajectory equation for the sphere (\ref{eqn:sphere_non_dim_01}) reduces to
\begin{equation}
        v'(t)
    =
    -
    \mathfrak{F}
    +
    2\mathfrak{M}
    A(t).
\end{equation}
In the quasi-static model, $A(t)$ is fully determined by the instantaneous $\delta_s$ at time $t$.

\section{Numerical implementation}\label{section:Numerical_implementation}
We introduce a homogeneous radial mesh with $n_{r}+1$ nodes and spacing $\delta_{r} = \mathfrak{L}/n_{r}$. We discretise time with an adaptive algorithm (detailed in sub-section \ref{section:iteration_on_geometry} below), sampling time at $n_t+1$ points. Moreover, we define the discrete approximations
\begin{equation}
    \eta_i^k \approx \eta(r_i,t_k),
    \quad u_i^k \approx u(r_i,t_k),
    \quad p_i^k \approx p(r_i,t_k),
    \quad h^k \approx h(t_k),
    \quad v^k \approx v(t_k),
\end{equation} 
for $i=1,\ldots,n_{r}+1$, and $k = 1,\ldots,n_t+1$; where $r_i = (i-1) \delta_{r} $ and $t_k = \sum_{l=1}^{k}\delta_t^k$, with $\delta_t^k$ being the $k$-th interval in our time mesh, and $\delta_t^1 \coloneqq 0$. 

The pressed area is resolved to the accuracy provided by the mesh. To that end, we introduce variable $q$, which takes the values of each possible number of contact points in our mesh (from $0$ to the total number of points under one radius). Variable $q$ represents the "candidate number of contact points"; and, when considering a given $q$, we assume that the boundary of the contact area is found exactly at the mid-point between nodes $q$ and $q+1$. 

The discrete formulation requires that we pose a different system of equations for each possible value of $q$; thus generating candidate solutions parameterised by $q$. We formulate the system for an arbitrary $q$ in what follows.

We define $\eta^{k+1,q}_i$, $u^{k+1,q}_i$, $p^{k+1,q}_i$, $h^{k+1,q}$ and $v^{k+1,q}$ as the candidate solutions associated to the assumption that there are exactly $q$ nodes in contact at time $t_{k+1}$; and we use the implicit Euler method in time and second order finite difference approximations in space. Hence, from system (\ref{eqn:eta_t})-(\ref{eqn:no_overlap}) we have
\begin{eqnarray}
    \eta_i^{k+1,q}-\delta_t^{k+1} u_i^{k+1,q} = \eta_i^k,
    &&
    u_i^{k+1,q} 
    -\delta_t^{k+1}
    \check{\kappa}_i^{k+1}(q)
    +\delta_t^{k+1}
    p_i^{k+1,q}
    =
    u^k_i
    -\delta_t^{k+1}\mathfrak{F},
\end{eqnarray}
for $k = 1, \ldots,n_t$ and $i = 1,\ldots,n_{r}$, where
\begin{equation}
    \label{eqn:check_kappa}
    \check{\kappa}_i^{k+1}(q)
    \coloneqq
    \left\{
    \begin{array}{cc}
        2 &  i \leq q,
        \\
        \frac{
        \eta_{i-1}^{k+1,q}
        -2\eta_i^{k+1,q}
        +\eta_{i+1}^{k+1,q}
        }{
        \left(\delta_{r}\right)^2}
        + \frac{
        \eta_{i+1}^{k+1,q}
        -\eta_{i-1}^{k+1,q}
        }{
        2(i-1) \left(\delta_{r}\right)^2
        } 
        & i>q,i>1,
        \\
        4
        \frac{
        \eta_{i+1}^{k+1,q}
        -\eta_i^{k+1,q}
        }{
        \left(\delta_{r}\right)^2}
        & q = 0,i=1;
    \end{array}\right.
\end{equation}

Moreover, for $k = 1,\ldots,n_t$ we have
\begin{eqnarray}
    h^{k+1,q}
    -\delta_t^{k+1}
    v^{k+1,q}
    =
    h^k,
    &&
    v^{k+1,q}
    -\delta_t^{k+1}\mathfrak{M} \mathcal{H}(q)p^{k+1,q}
    =
    -\delta_t^{k+1}\mathfrak{F},
\end{eqnarray}
where $p^{k+1,q} = [p^{k+1,q}_1, p^{k+1,q}_2,\ldots,p^{k+1,q}_{n_{r}}]^T$ and $\mathcal{H}(q)$ is the integral operator, represented by a row vector, that interpolates the radial direction using the trapezium rule for integration in $[0,r_q]$ and between $[r_q,(r_q+r_{q+1})/2]$, with $p = 0$ for all $r \geq \left(r_q+r_{q+1}\right)/2$.

Furthermore; for $k=0,1,\ldots,n_t$ we have
\begin{eqnarray}
    \eta^{k+1,q}_{n_{r}+1} = 0,
    &&
    u^{k+1,q}_{n_{r}+1} = 0;
\end{eqnarray}
for $i = 1,\ldots,n_{r}+1$, we have
\begin{eqnarray}
    u^1_{i} = 0,
    &&
    \check{\kappa}_i^1(q=0) = \mathfrak{F};
\end{eqnarray}
for $i>q$, we have
\begin{eqnarray}
    p_i^{k+1,q} = 0.
\end{eqnarray}
We also have 
\begin{eqnarray}
    h^1 = 1+\eta_1^1,
    &&
    v^1 = -\mathfrak{U};
\end{eqnarray}
and, for $i\leq q$, we have
\begin{eqnarray}
    \eta_i^{k+1,q}=h^{k+1,q}_i+s_i,
    &&
    u_i^{k+1,q} = v^{k+1,q},
\end{eqnarray}
with 
\begin{equation}
    s_i= s(r_i), \text{ for }i = 1,\ldots,q_{\text{max}}
\end{equation}
where $q_{\text{max}}$ is $1$ plus the integer part of $1/\delta_{r}$. Also, we have for $q<i\leq q_{\text{max}}$
\begin{eqnarray}
    \label{eqn:no_overlap_discrete}
    \eta_i^{k+1,q}<h^{k+1,q}+s_i.
\end{eqnarray}

Finally, we define $m^{k+1}$, i.e. the number of nodes in contact at time $t_{k+1}$, as
\begin{equation}
    \label{eqn:tangency_argmin}
    m^{k+1} \coloneqq \text{argmin}_q
    \left|
    \sigma_{q+\frac{1}{2}}
    -
    \frac{
    \eta_{q+1}^{k+1,q}
    -\eta_q^{k+1,q}
    }{
    \delta_{r}
    }
    \right|,
\end{equation}
with 
\begin{equation}
    \label{eqn:tangency_discrete}
    \sigma_{i+\frac{1}{2}}
    =
    s'\left(\frac{r_i+r_{i+1}}{2}\right);
\end{equation}
as well as
\begin{equation}
   \eta_{i}^{k+1} = \eta_{i}^{k+1,q}|_{q = m^{k+1}},
   \qquad 
   u_{i}^{k+1} = u_{i}^{k+1,q}|_{q = m^{k+1}},
   \qquad
   p_{i}^{k+1} = p_{i}^{k+1,q}|_{q = m^{k+1}},
\end{equation}
for $i=1,2,\ldots,n_r+1$, and
\begin{equation}
   h^{k+1} = h^{k+1,q}|_{q = m^{k+1}},
   \qquad 
   v^{k+1} = v^{k+1,q}|_{q = m^{k+1}}.
\end{equation}

At each time step, $m^{k+1}$ is to be found using an iterative method, detailed in subsection \ref{section:iteration_on_geometry} below. We note that, on a non-moving mesh, condition (\ref{eqn:tangency_discrete}) can only be satisfied to the accuracy of $\delta_{r}$. An approach based on the finite element method, using the spine method for moving meshes, is being developed and will be detailed in a separate article.

\subsection{System matrices}
The discrete version of our impact problem can be summarised as follows
\scriptsize
\begin{equation}
    \left[
    \begin{array}{ccccc}
        \mathcal{I} 
        & -\delta_t^{k+1}\mathcal{I}
        & 0
        & 0
        & 0
        \\
        \delta_t^{k+1}
        \mathcal{A}(q)
        & \mathcal{I}
        & \delta_t^{k+1}\mathcal{I}
        & 0
        & 0
        \\
        0
        & 0
        & 0
        & 1
        & -\delta_t^{k+1}
        \\
        0
        & 0
        & -\delta_t^{k+1}\mathcal{H}(q)
        & 0
        & 1
    \end{array}
    \right] 
    \left[
    \begin{array}{c}
        \eta^{k+1,q}
        \\
        u^{k+1,q}
        \\
        p^{k+1,q}
        \\
        h^{k+1,q}
        \\
        v^{k+1,q}
    \end{array}
    \right]
    = 
    \left[
    \begin{array}{c}
        \eta^{k}
        \\
        u^{k}
        \\
        h^{k}
        \\
        v^{k}
    \end{array}
    \right]
    +
    \left[
    \begin{array}{c}
        0
        \\
        \mathcal{U}(q)
        \\
        0
        \\
        -
    \delta_t^{k+1}\mathfrak{F}
    \end{array}
    \right],
\end{equation}
\normalsize
where $\eta^k = [\eta_1^k,\eta_2^k,\ldots,\eta^k_{n_r}]^T$ and $u^k$, $\eta^{k+1,q}$, $u^{k+1,q}$, $p^{k+1,q}$ are defined analogously. Moreover, $\mathcal{I}$ is the identity matrix of size $n_{r}\times n_{r}$, $\mathcal{A}(q)_{i,j} = 0$ if $i\leq q$, and $\sum_jA(q)_{i,j}\eta^{k+1,q}_j = \check{\kappa}_i^{k+1}(q)$, otherwise. Moreover, \begin{equation} 
    \mathcal{U}(q) = \delta_t^{k+1}\left(\mathcal{V}(q)-
\mathfrak{F}\mathcal{R}\right),
\end{equation}
where $\mathcal{R}$ is a column vector of ones with $n_{r}$ entries, and $\mathcal{V}_i(q) = 2$ if $i\leq q$, and $\mathcal{V}_i(q) = 0$, otherwise. 

The system above is rectangular of size $(2n_{r}+2)\times(3n_{r}+2)$, but it can be reduced to a square system of size $(2n_{r}-q+2)\times(2n_{r}-q+2)$ by imposing our constraints. To make the process of implementing constraints as transparent as possible, we introduce the following notation.

Given matrix $\mathcal{B}$, we define $\left[\mathcal{B}\right]^q$ as the matrix composed of the first $q$ columns of $\mathcal{B}$, and $\left[\mathcal{B}\right]^{q'}$ as the matrix formed by all columns of $\mathcal{B}$, except for the first $q$. An entirely analogue definition is done with rows of $\mathcal{B}$, and sub-indexes, so that
\begin{equation}
    \mathcal{B}
    =
    \left[ 
    \begin{array}{cc}
        \left[\mathcal{B}\right]^q
        & 
        \left[\mathcal{B}\right]^{q'}
    \end{array}
    \right] 
    =
    \left[ 
    \begin{array}{cc}
        \left[\mathcal{B}\right]_q 
        \\
        \left[\mathcal{B}\right]_{q'}
    \end{array}
    \right]
    =
    \left[ 
    \begin{array}{cc}
        \left[\mathcal{B}\right]^q_q 
        &
        \left[\mathcal{B}\right]^{q'}_{q} 
        \\
        \left[\mathcal{B}\right]^q_{q'} 
        & 
        \left[\mathcal{B}\right]^{q'}_{q'}
    \end{array}
    \right],
\end{equation}
with $\left[\mathcal{B}\right]^q_q \coloneqq \left[\left[\mathcal{B}\right]^q\right]_q$ and analogously for the other three blocks in the rightmost matrix above.

Applying the constraints of the problem we have
\scriptsize
\begin{equation}
    \left[
    \begin{array}{ccccc}
        \left[\mathcal{I}\right]^{q'}_{q'} 
        & -\delta_t^{k+1}\left[\mathcal{I}\right]^{q'}_{q'} 
        & 0
        & 0
        & 0
        \\
        \\
        \delta_t^{k+1}
        \left[\mathcal{A} (q)\right]^{q'}
        & \left[\mathcal{I}\right]^{q'} 
        & \delta_t^{k+1}\left[\mathcal{I}\right]^q
        & \mathcal{X}
        & \mathcal{Y}
        \\
        \\
        0
        & 0
        & 0
        & 1
        & -\delta_t^{k+1}
        \\
        \\
        0
        & 0
        & -\delta_t^{k+1}\left[\mathcal{H}(q)\right]^{q}
        & 0
        & 1
    \end{array}
    \right] 
    \left[
    \begin{array}{c}
        \left[\eta^{k+1,q}\right]_{q'}
        \\
        \\
        \left[u^{k+1,q}\right]_{q'}
        \\
        \\
        \left[p^{k+1,q}\right]_q
        \\
        \\
        h^{k+1,q}
        \\
        \\
        v^{k+1,q}
    \end{array}
    \right]
    = 
    \left[
    \begin{array}{c}
        \left[\eta^{k}\right]_{q'}
        \\
        \\
        u^{k}
        +
        \mathcal{W}
        \\
        \\
        h^{k}
        \\
        \\
        \beta^k
    \end{array}
    \right],
\end{equation}
\normalsize
where
\begin{eqnarray}
    \mathcal{X}
    =
    \delta_t^{k+1}
    \left[\mathcal{A}(q)\right]^q
    \left[\mathcal{R}\right]_q,
    &&
    \mathcal{Y} = 
    \left[
    \begin{array}{c}
        \left[\mathcal{R}\right]_q 
        \\0
    \end{array} 
    \right],
\end{eqnarray}
\begin{eqnarray}
    \mathcal{W}
    =
    \mathcal{U}(q)
    +\left[\mathcal{A}(q)\right]^q\left[\mathcal{S}\right]_q
\end{eqnarray}
with 
\begin{equation}
    \mathcal{S} = [s_1,s_2,\ldots,s_{q_{\text{max}}}]^T,
\end{equation}
and
\begin{equation}
    \beta^k=v^{k}-\delta_t^{k+1}\mathfrak{F}.
\end{equation}

The manipulations above, transform the problem of finding the solution of the next time step to the solution of a square linear system of the form 
\begin{equation}
\label{eqn:system_of_q_and_dt}
    \mathcal{M}\left(q,\delta_t^{k+1}\right)x^{k+1}(q)
    =
    b^k\left(q,\delta_t^{k+1}\right),
\end{equation}
with $2n_{r}-q+2$ unknowns, under the assumption that the correct contact area for that time step contains exactly $q$ points. However, $q$ itself is an unknown, and it will be found using an iterative method. We highlight that once we solve system (\ref{eqn:system_of_q_and_dt}) for a given candidate number of contact points $q$, conditions (\ref{eqn:no_overlap_discrete}) and (\ref{eqn:tangency_argmin}) are yet to be verified. These two conditions will be checked to determine which value of $q$ is assigned to $m^{k+1}$, using equation (\ref{eqn:tangency_argmin}) and following the method presented in sub-section \ref{section:iteration_on_geometry} below.

\subsection{An iteration on the geometry}
\label{section:iteration_on_geometry}
Our assumption that the contact surface $S(t)$ changes continuously is reflected, in our discrete approximation, by the condition that the boundary of the contact area can move by at most one interval of the spatial mesh per time step. To properly impose the condition, we must be able to reduce the time step whenever this is needed to capture the velocity of the boundary of the contact area. 

The continuous dependence of the location of $C(t)$ on time implies that, at any given time, we only need to look for the location of the boundary in the vicinity of its previous location. Consequently, our implementation, described in Appendix \ref{sec:Algorithm}, is based on finding the closest local minimum in tangency error (see equation \ref{eqn:tangency_argmin}).

In practice, we only test up to five points (the previous number of contact nodes plus and minus one and two nodes) before we decide if a reduction of the time step is needed. Once we have calculated the tangency error for each of these five configurations (or less in some cases) we can determine whether the local minimum in tangency error is at most $\delta_r$ away from the location of the boundary of the contact area at the previous time step. If this is the case, we accept that solution as the best approximation that our non-moving mesh can provide for the location of $L(t)$, and assign the value of $m^{k+1}$, accordingly. Alternatively, if the local minimum is found two points away from the previous location of the boundary, we halve the time step and repeat the procedure. 

The algorithm implemented enlarges the time step when the need for an extra fine step is overcome. This is achieved by testing larger time steps when these do not need to be reduced. Nevertheless, the algorithm also includes restrictions to prevent unchecked growth of time steps. To this end, two additional conditions are imposed to control the rate at which the time step increases, there is a maximum allowed time-step size, explicitly prescribed, and there is a condition to only allow a time step to be twice the prior time step at most. The latter condition is mainly meant to have a relatively regular sampling in time. \newline Furthermore, we ensure that every integer multiple of the maximum allowed time step is used as one of the discrete times in our algorithm. This allows the use of regular time samples for visualisation of results.

The procedure described relies on our finding that the tangency error $s'(r_c(t))-\partial_{r}\eta(r_c(t),t)$ behaves monotonically in all cases tested. In particular, it is always found to be positive for pressed areas that are larger than the optimal, and negative for pressed areas that are smaller than optimal; replicating the situation found in the case of impacts on the free surface of a fluid bath (see figure 2 in \cite{Galeano-RiosEtAl2017}).

In the simulations presented in what follows, the spacing of the radial mesh and the maximum allowed time step are chosen so that halving either yields a difference of less than one percent in contact time and maximum surface deflection. This was achieved by setting $\delta_{r}$ numerically in such a way that, in dimensionless units, $\delta_{t} \leq \delta_{r} \leq 5 \times 10^{-3} $. A github repository with all the code needed to replicate our results was made available at \\ \href{https://github.com/elvispy/kinematic-match-sphere}{https://github.com/elvispy/kinematic-match-sphere}.

\section{Experiments}\label{section:Experiments}

\begin{figure}
    \centering
    \includegraphics*[width =1\textwidth]{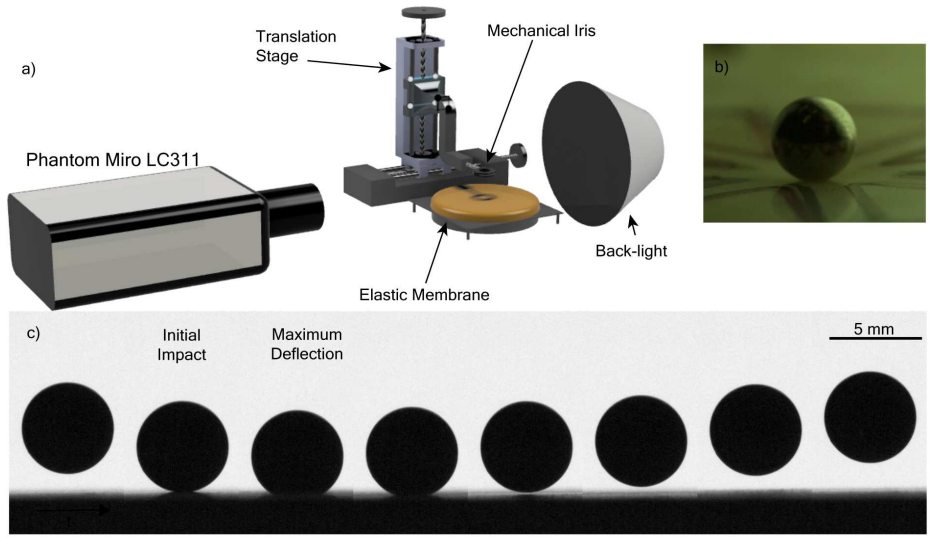}
    \caption{(a) A 3D rendering of the experimental setup. (b) Ceramic sphere resting on a tensioned membrane. (c) A sequence of images depicting the  initial stages of impact and subsequent bounce for a ceramic sphere of diameter $D= 4.73 \ \mathrm{ mm}$. The time interval between images is 1 ms. }
    \label{fig:ExpSetup}
\end{figure}
A rendering of the experimental set-up is depicted in figure \ref{fig:ExpSetup}. In each trial, spheres were dropped from a mechanical iris that is connected to a 2 degree-of-freedom linear stage that allows for precise and repeatable release of the spheres. The elastic membrane is clamped to a square holding plate with circular cut-out, which is then stretched over a hollow vertical cylinder of mean diameter $105$ mm. The membranes used in these experiments are HYTONE LS-034 natural rubber latex sheets of thickness $0.3$ mm and have a material density of $0.98 \text{ } \mathrm{ g}/\mathrm{cm}^{3}$. The top edge of the cylinder is rounded with a $5$ mm radius to ensure smooth contact with the membrane. The vertical cylinder can be precisely levelled by adjusting three levelling spring supports.  The membrane holding plate is then securely fastened to an optical table. The vibration isolation provided by the optical table ensured minimal disturbances on the membrane prior to impact. A Phantom Miro LC311 camera with a Nikon Micro 200 mm lens was used for the
video capture. The camera was mounted directly on the optical table with the back edge of the camera elevated slightly for a downward viewing angle of $\sim5^{\circ}$. Images were captured at 10,000 frames per second with an exposure time of 99.6  $\mathrm{\mu s}$. Two spheres of diameter $4.73$ mm and different densities were used in this study: one of SAE 304 stainless steel $\rho_s = 7.93 \ \text{g}/\mathrm{cm}^{3}$ and the other of Silicon Nitride ceramic $\rho_s = 3.25 \ \text{g}/ \mathrm{cm}^{3}$.  Release heights were varied to achieve impact velocities from $25$ to $100$ $\mathrm{cm}/\text{s}$ .

Spheres were released from the mechanical iris at a range of heights, beginning at approximately one sphere diameter above the membrane. To characterise error, a minimum of 5 trials were completed at each height, and spheres were routinely cleaned using isopropyl alcohol and dried before being re-used in the experiments. After each increase in height, the membrane was wiped using dust-free optical lens cleaning paper. The raw video data was processed using a custom code written in MATLAB that uses a Canny edge detection. The top and bottom edges in the image corresponding to the north and south poles of the sphere, respectively, were then recorded. Initial contact ($t=0$) was determined as the time where the actual sphere and its reflection in the membrane first met. Due to the slight downward angle of the camera toward the membrane, this instant was resolved in all trials. During contact, the south pole was obscured by the membrane edge, and the trajectory of the south pole of the sphere was determined by shifting the top trajectory down by one sphere diameter. For the range of impact speeds tested, the top point on the sphere was resolvable for all times during contact. 

To determine the membrane tension, we placed a large solid stainless steel sphere of radius $R=15.875$ mm at the centre of the membrane and measured the maximum static displacement $\delta_s$.  
To relate these to the membrane tension $\tau$, we balance the vertical forces on the sphere at equilibrium using the static membrane solution outlined in \S\ref{section:Problem_formulation}\ref{sec:qs} and rearrange to get
\begin{align}
    \tau = \frac{2\rho_s R^2 g}{3r_c^2}. 
    \label{tension}
\end{align}
In summary, we measured $\delta_s$ from a still camera image, then solved for $r_c$ numerically in equation \eqref{delta} and use equation \eqref{tension} to determine the tension. Additionally, we compared the solution of the linearised problem above to the solution including the fully nonlinear curvature term (which yields a catenoid solution\cite{GiletAndBush2009})  and found negligible quantitative differences for our current \newline 
experimental parameters.

In the present work, contact time, $t_c$, is defined as the time duration from when the bottom of the sphere touches the membrane to the time the bottom of the sphere returns to that height. Due to the nature of visualisation set-up, it was impossible to accurately determine when the spheres lost physical contact with the membrane. Each bounce is also characterised by its coefficient of restitution, $\alpha$, which is defined here as the negative of the normal exit velocity, $V_e$, divided by the normal impact velocity,
$V_0$. The exit velocity is taken to be the velocity of the top of the sphere measured exactly at the contact time, $t_c$. $V_e$ and $V_0$ are determined by fitting a quadratic polynomial to both the incoming and outgoing trajectories in MATLAB, ensuring that at least 30 data points (frames) were used in each fit to minimise error. 
Additionally, we measure the maximum membrane deflection $\delta $ as the lowest point in the bottom trajectory of the sphere. Error bars are quantified as the standard deviation of the respective measurement over at least 5 experimental trials. 

\section{Results}\label{section:Results}
\begin{figure}
    \centering
    \begin{tabular}{cc}
    \includegraphics*[width = .35\textwidth]{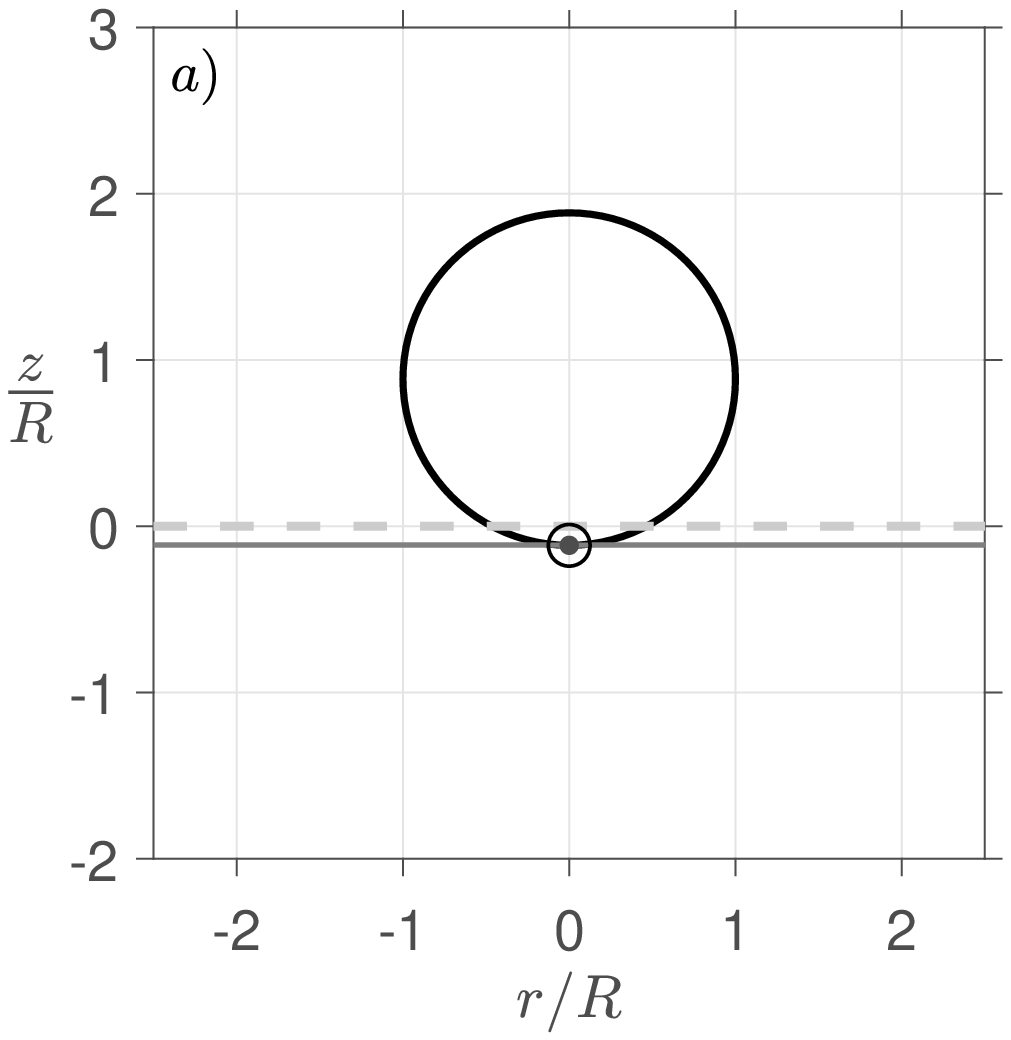} &  
    \includegraphics*[width = .35\textwidth]{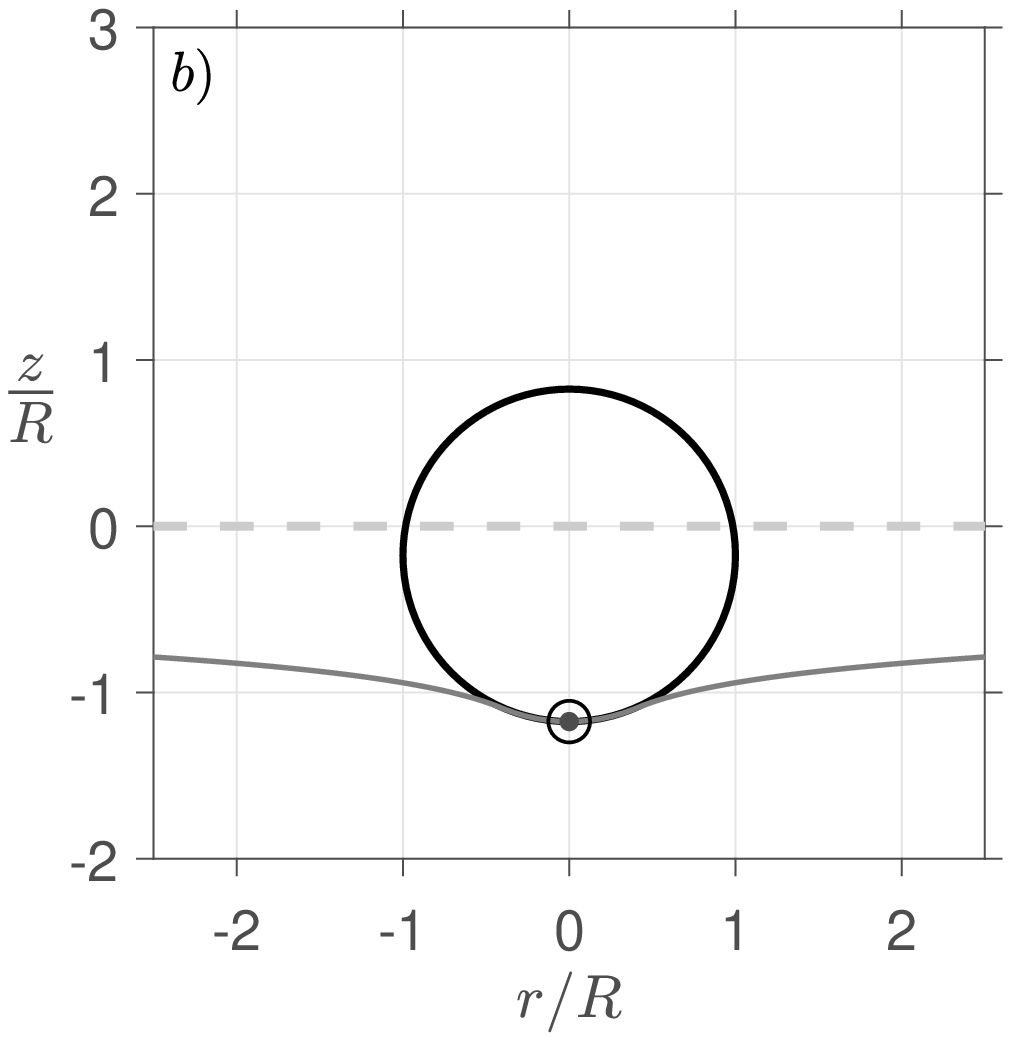}
    \\
    \includegraphics*[width = .35\textwidth]{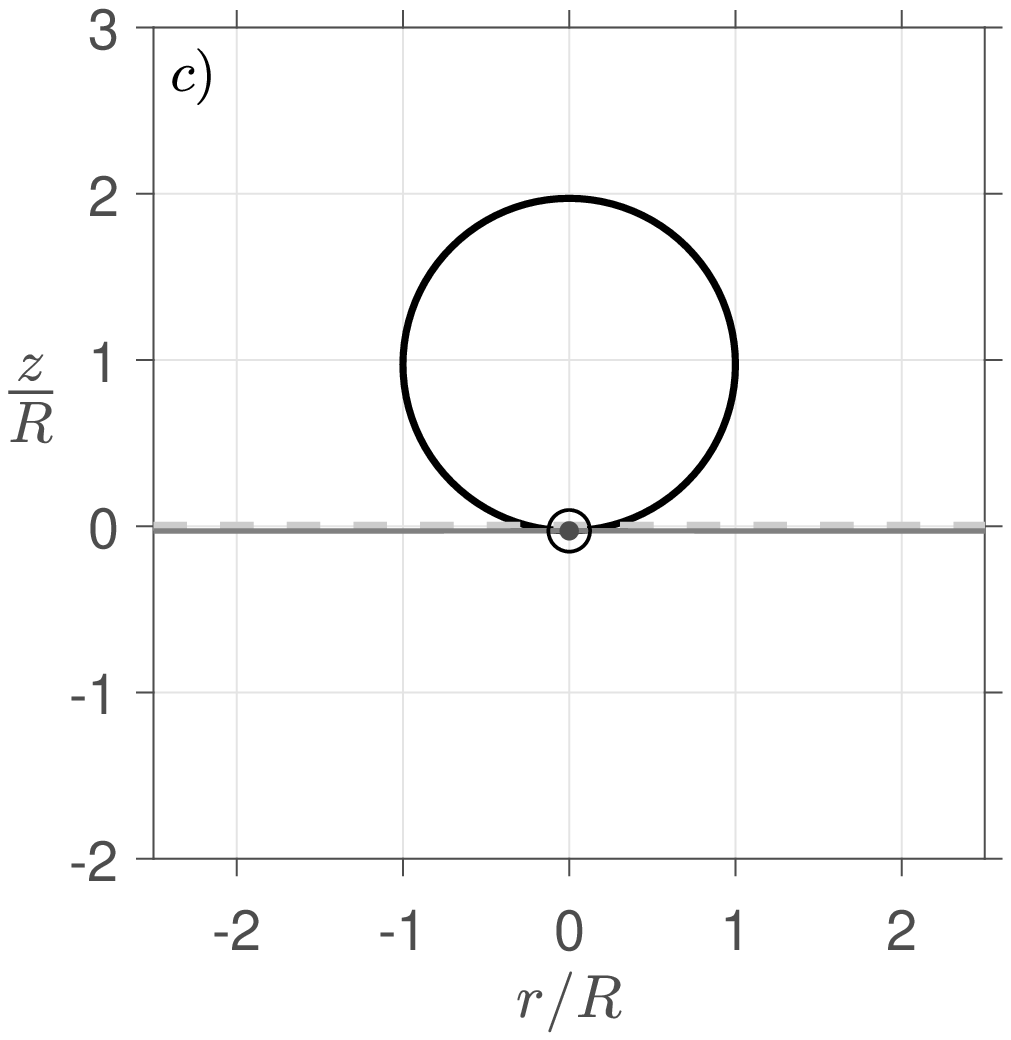}
    & 
    \includegraphics*[width = .35\textwidth]{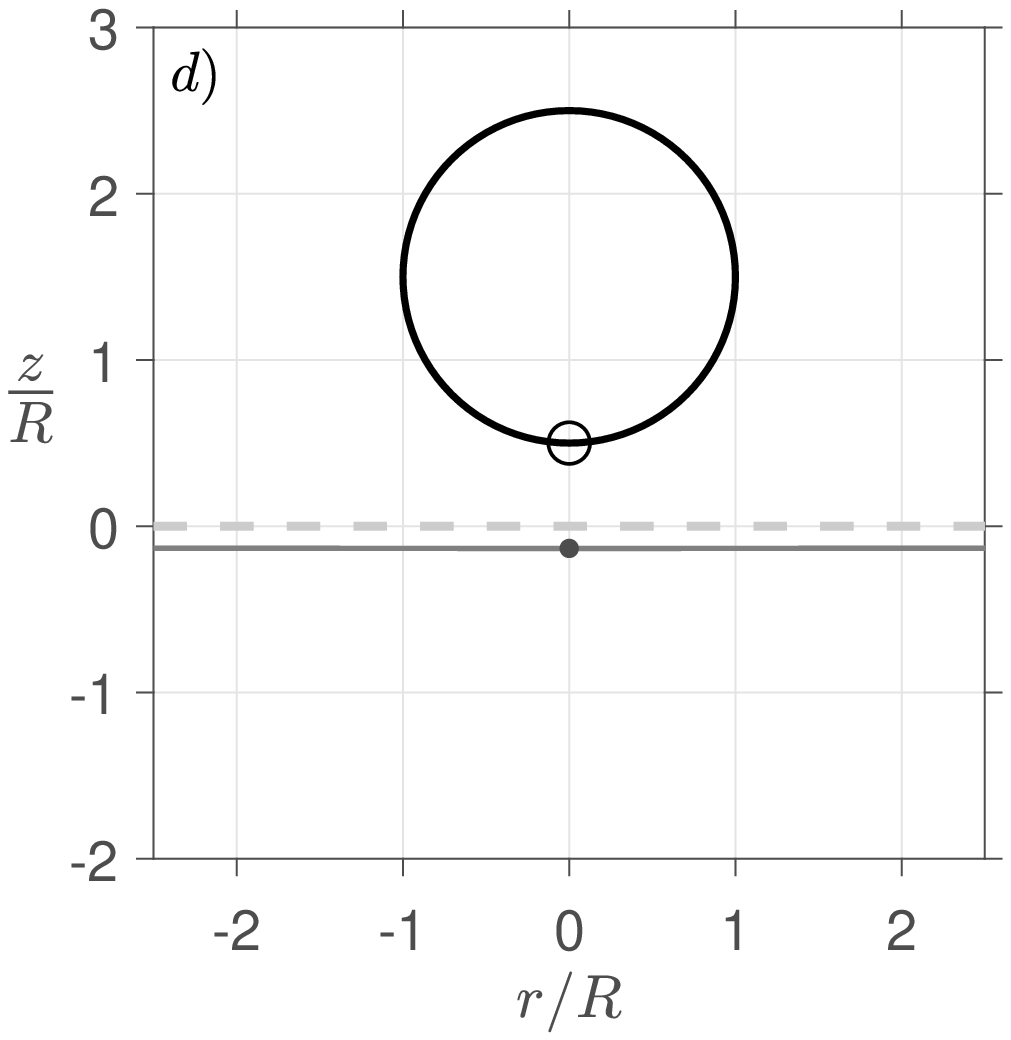}
    \end{tabular}
    % rS = 1, R_f = 10, Tm = 1, v_k = -0.15
    \caption{Simulation of impact and rebound for $\mathfrak{F} = 1.648 \times 10^{-4} $, $\mathfrak{M} = 5.142 \times 10^{-4}$, $\mathfrak{L} = 10 $  and $\mathfrak{U}=-1.944\times 10^{-2}$. The trajectories of the centre of the membrane (\raisebox{.5pt}{{\tiny \textcolor{mygrey}{$\CIRCLE$}}}) and the lowest point of the sphere ($\ocircle$) are used to define the contact time, maximum deflection and coefficient of restitution of the impact. The dashed line corresponds to the $z=0$ plane. Panels correspond to: $a$) imminent impact at $t=0$, $b$) maximum absolute deflection, $c$) imminent take-off, and $d$) flight following rebound.}
    \label{fig:Example}
\end{figure}

Our simulations show the sphere landing on the membrane, deforming it as the pressed surface expands, and bouncing back as the pressed surface contracts and then vanishes (see figure \ref{fig:Example}). Simulations are run until the centre of the membrane starts to move downward, following lift-off. However, the method is able to capture repeated bounces, as shown in a video animation of these results, which is made available as supplementary material. We follow \cite{Galeano-RiosEtAl2021} and we check that all simulations satisfy the condition $|\nabla\eta(r,t)|<1$, throughout the simulation, as a consistency check for our linear approximation of curvature outside the pressed area.

To facilitate comparisons with the experimental results, we measure $t_c$, $\delta$ and $\alpha$ (as defined in the experiments). However, it should be noted that there is no difficulty in obtaining the exact time at which the sphere detaches from the membrane in our simulations, therefore it is also possible to use such instant as the basis for the definition of contact time and coefficient of restitution, if needed.

\subsection{Comparisons to experiments}
\begin{figure}
    \centering
    \begin{tabular}{cc}
    \hbox{\hspace{0.25em}\includegraphics[width = .395\textwidth]{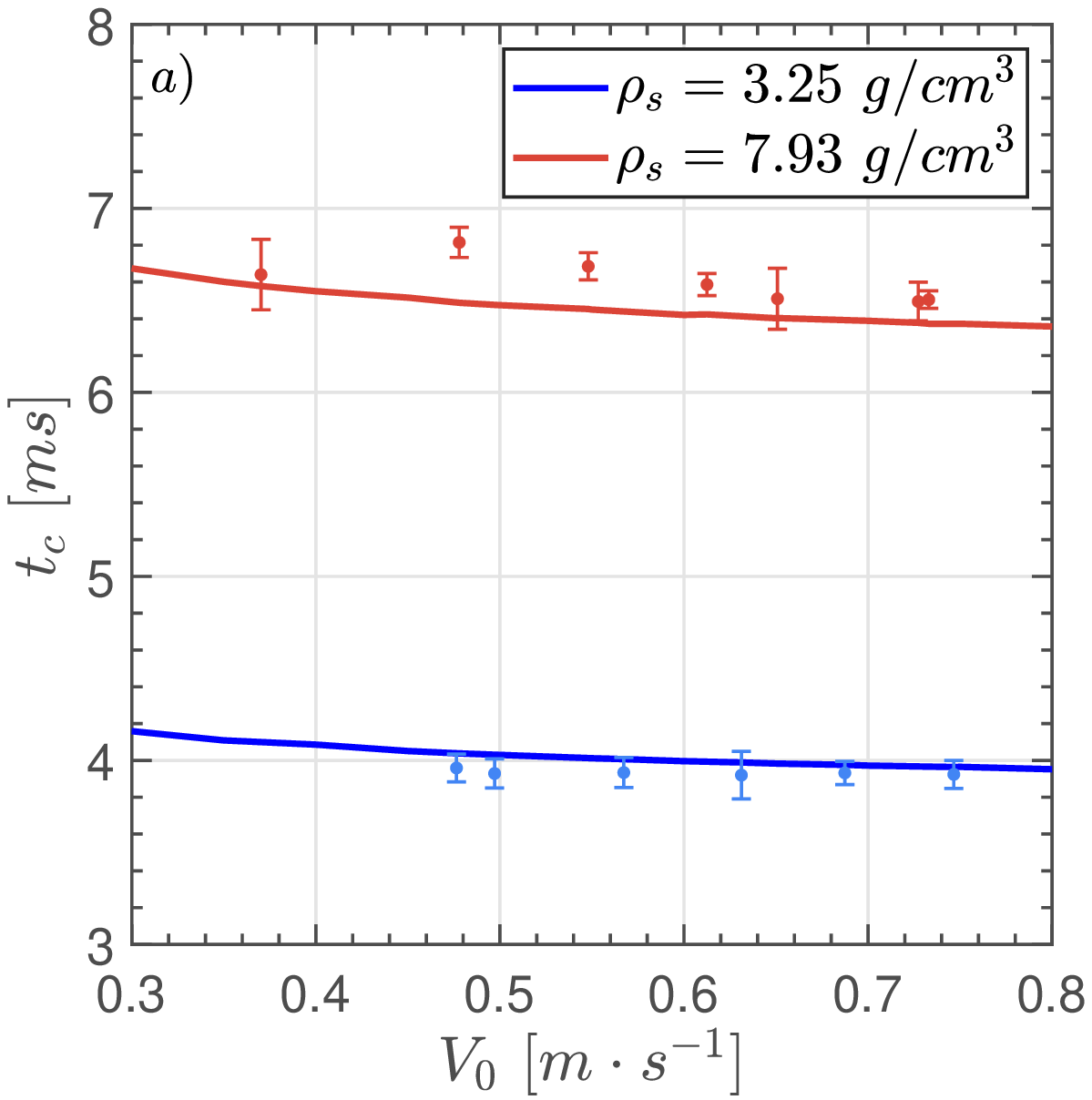}}
         & 
\includegraphics[width = .405\textwidth]{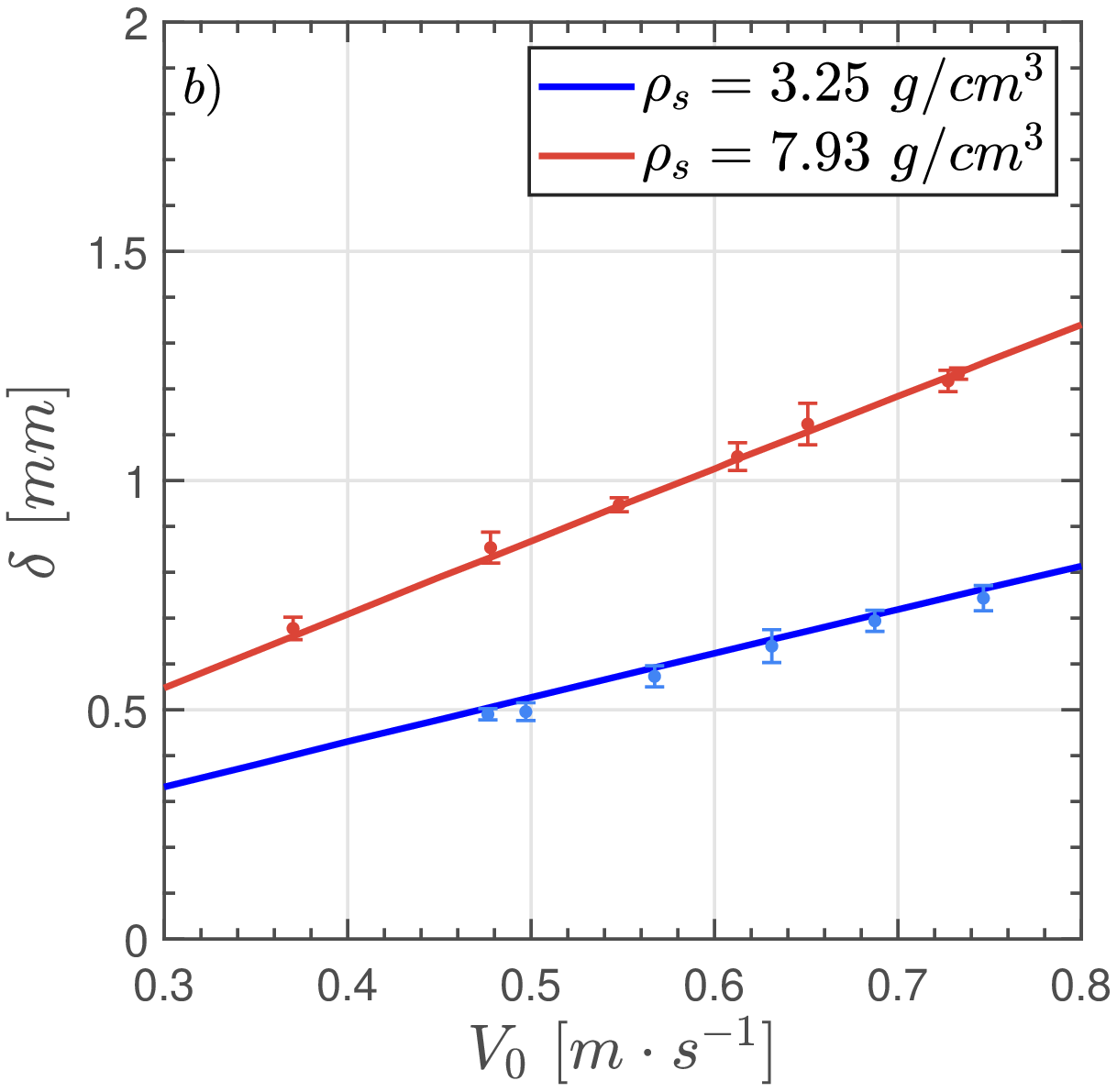} \\
    \includegraphics[width = .4\textwidth]{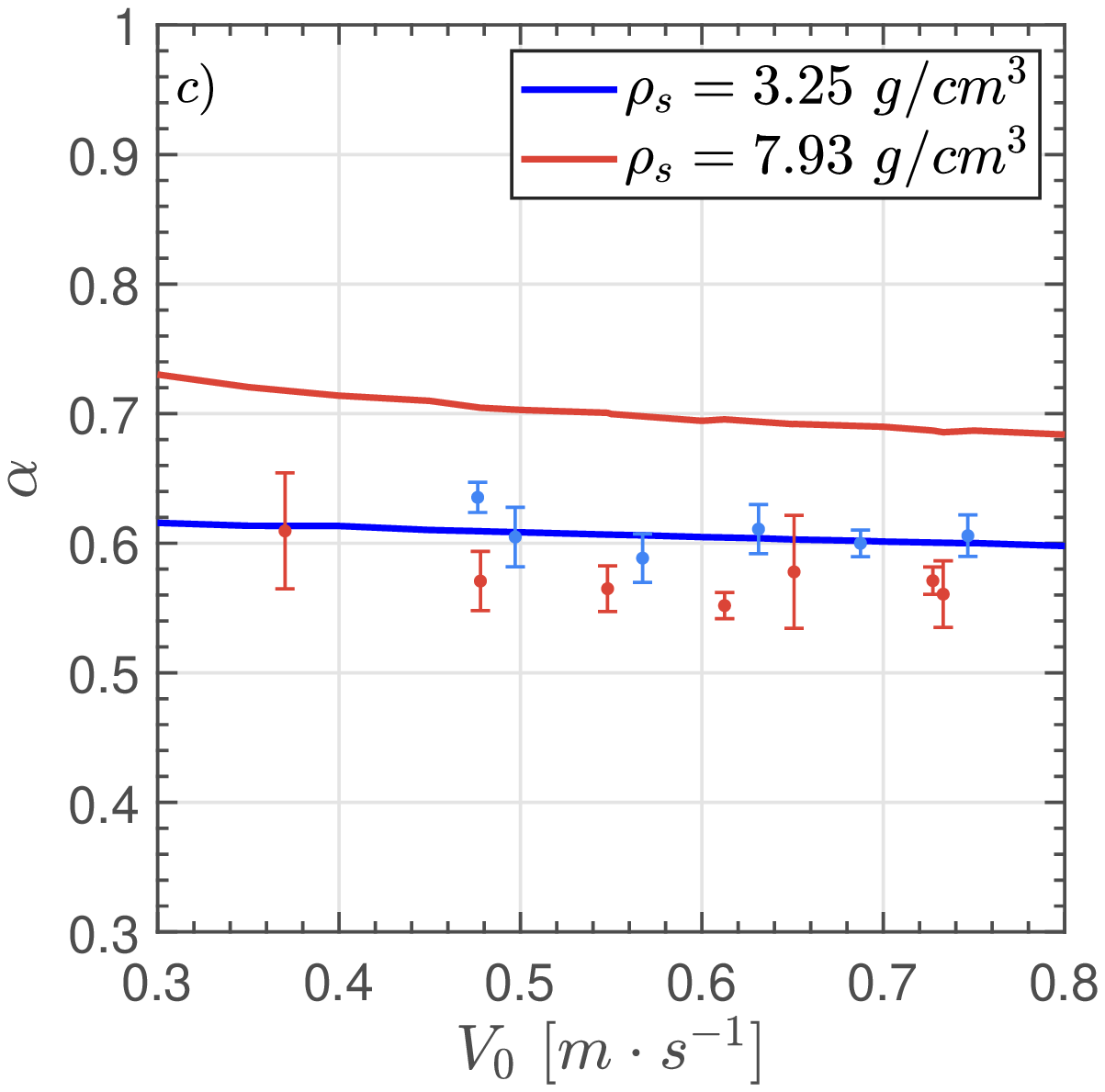}
         & 
    \includegraphics[width = .409\textwidth]{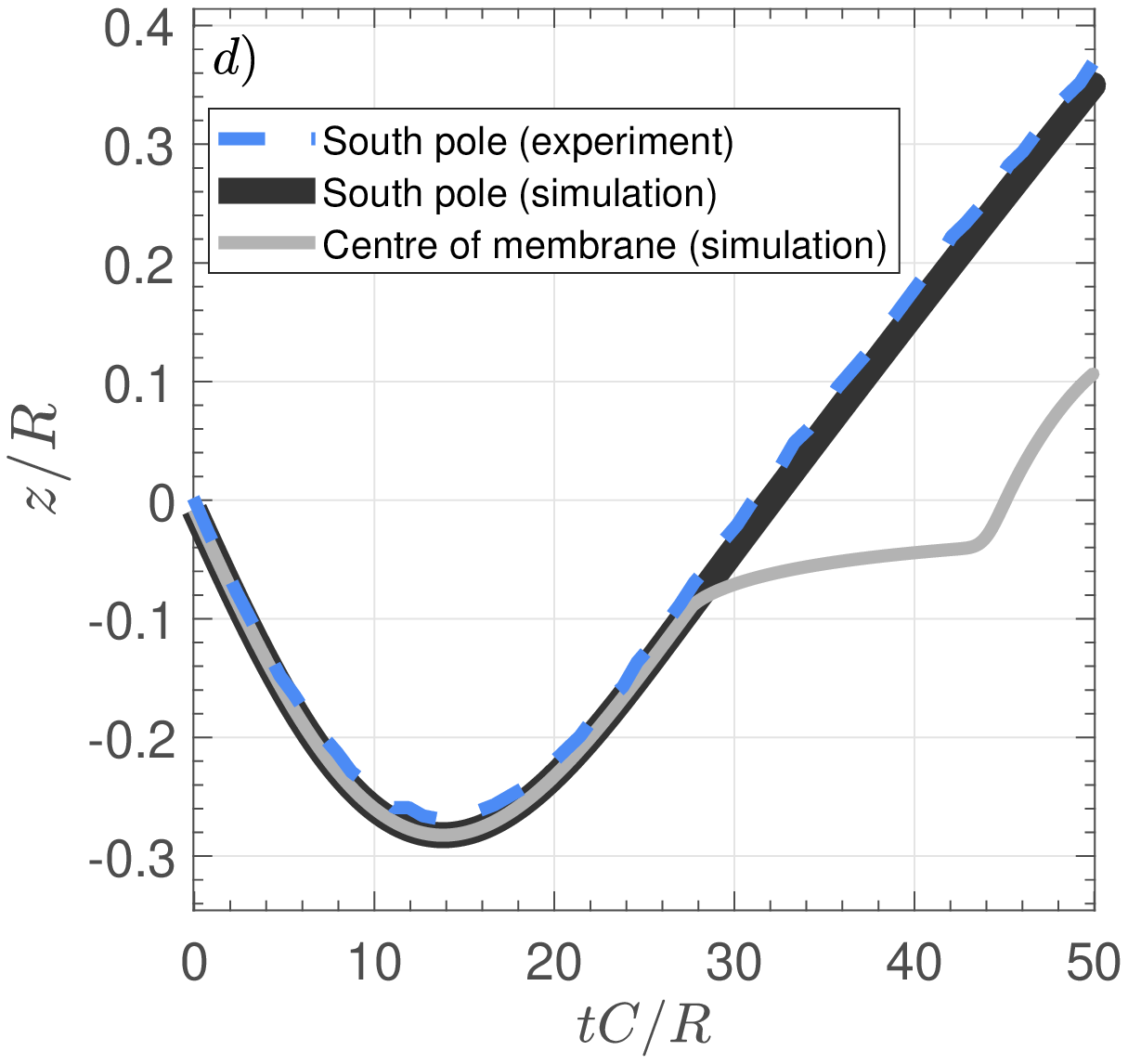}
    
    \end{tabular}
    \caption{Comparison of full simulation predictions (solid lines), quasi-static model predictions (dotted lines), and experimental measurements for contact time (a), maximum surface deflection (b), coefficient of restitution (c), and south pole trajectory (d) for $\Lambda = 52.5$ mm, $R = 2.38$ mm, $\mu = 0.3$ kg/m$^2$ and $\tau = 107$ N/m (i.e. $\mathfrak{L} = 22.06$, $\mathfrak{F} = 6.54 \times 10^{-5}$), for both $\rho_s = 3.25$ g/cm$^3$ (blue lines and markers) and $\rho_s = 7.93$ g/cm$^3$ (red), (i.e. $\mathfrak{M} = 9.26 \times 10^{-3}$ and $\mathfrak{M} = 3.79 \times 10^{-3})$. Experimental values are shown with error bars using the same colour coding as solid lines that represent model predictions. Trajectories of the south pole are compared for $\mathfrak{L} = 22.06$, $\mathfrak{F} = 6.54 \times 10^{-5}$, $\mathfrak{M} = 9.26 \times 10^{-3}$, and $\mathfrak{U} = 3.34 \times 10^{-2}$.
    }\label{fig:Comparison}
\end{figure}

We compare our full simulation and corresponding quasi-static model predictions to our experimental results for the set-up described in section \ref{section:Experiments} using two different sphere densities over a range of impact velocities. In our experiments, the non-dimensional quantity $m/(\mu \Lambda^2)$ defined in equation (\ref{eqn:massratio}) takes a value 0.22-0.54, signifying that we are outside of the quasi-static regime for the parameters considered here. Our predictions for contact time $t_c$ and maximum surface deflection $\delta$ are in line with our experimental results for both sphere densities used, and our predictions for the coefficient of restitution $\alpha$ match the experiments for the lower sphere density case, as can be seen in figure \ref{fig:Comparison}. Agreement in the coefficient of restitution is not equally good for the larger density sphere. This error in the coefficient of restitution is, to some extent, expected in the case of heavier spheres; in which the resulting larger deformation may mean that dissipation mechanisms and material nonlinearities, not considered in the present model, are of importance to the rebound. The quasi-static model underpredicts the contact time and overpredicts the maximum surface deflection for the cases studied in figure \ref{fig:Comparison}. Furthermore, we measure $\alpha<1$ indicative of energy transfer to the membrane during impact, an effect that is captured by the full model but not the quasi-static model. The general trends in our data and predictions, specifically the near independence of the contact time and coefficient of restitution with the impact velocity and the approximate linear relationship between the maximum deformation and impact velocity, are consistent with classical predictions of the rebound of a linear mass-spring-damper model under weak gravity \cite{nagurka2004mass}.

The agreement of our predictions with the experiments is not limited to the metrics mentioned above, the full trajectory is also well predicted by our method. In panel $d$ of figure \ref{fig:Comparison}, we compare the prediction for the trajectory of the "south pole" of the sphere with the experimental measurement of the trajectory for the same physical parameters. The corresponding prediction of the quasi-static model is also shown, with poorer agreement to the measured trajectory.  In particular, the quasi-static model is unable to capture the asymmetry between the incoming and outgoing segments of the trajectory. Videos of an experiment and an animation of the simulation results for this bounce are made available as part of the supplementary material. 

We also attempted a comparison of our model predictions with the results reported in \cite{CourbinEtAl2006}. Unfortunately, a direct comparison was impossible, as the membrane tension used for each bounce was not reported. Instead, \cite{CourbinEtAl2006} reports a range of tensions used in their experiments. Given the information provided, the best that we could do was to test whether our predictions for that range of tensions was in line with their results. Indeed, our results for the minimum and maximum tensions reported in \cite{CourbinEtAl2006} produce an interval of possible values for the maximum deflection and the contact time that is consistent with the experimental results obtained in \cite{CourbinEtAl2006} for the lighter spheres used in that work.  Coefficients of restitution were not reported in \cite{CourbinEtAl2006}.

\subsection{Further findings}
One benefit of the present model is that it allows us to obtain detailed predictions for the evolution of variables such as the pressure distribution, which is more difficult to measure experimentally. Moreover, our model enables us to explore regimes that are challenging to experiment on, such as the low $\mathfrak{U}$ limit; in which, incidentally, our modelling assumptions are more readily satisfied.

\subsubsection{Coefficient of restitution}
An important quality of the method here considered lies in the fact that it captures the mechanism by which waves are generated over the contact. There are several useful implications of this virtue of the KM. In particular, we are able to estimate the transfer of energy to the impacted surfaces. This is reflected, for example, in the possibility to successfully predict the coefficient of restitution that results from the impact of a rigid solid onto a complex substrate, as was shown in \cite{Galeano-RiosEtAl2021}.

\begin{figure}
    \centering
    \begin{tabular}{cc}
    \includegraphics[width = .39\textwidth]{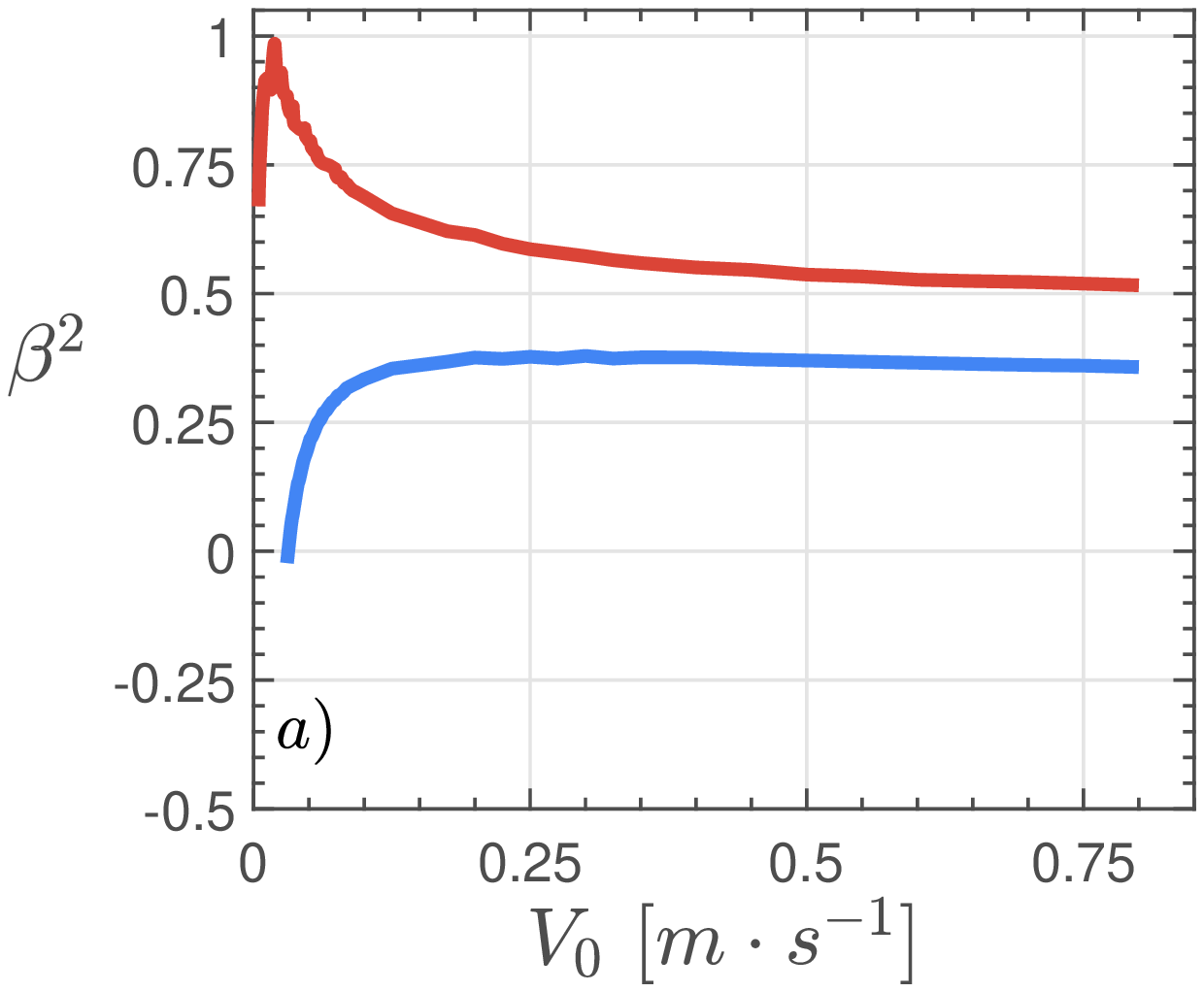}
         & 
    \includegraphics[width = .56\textwidth]{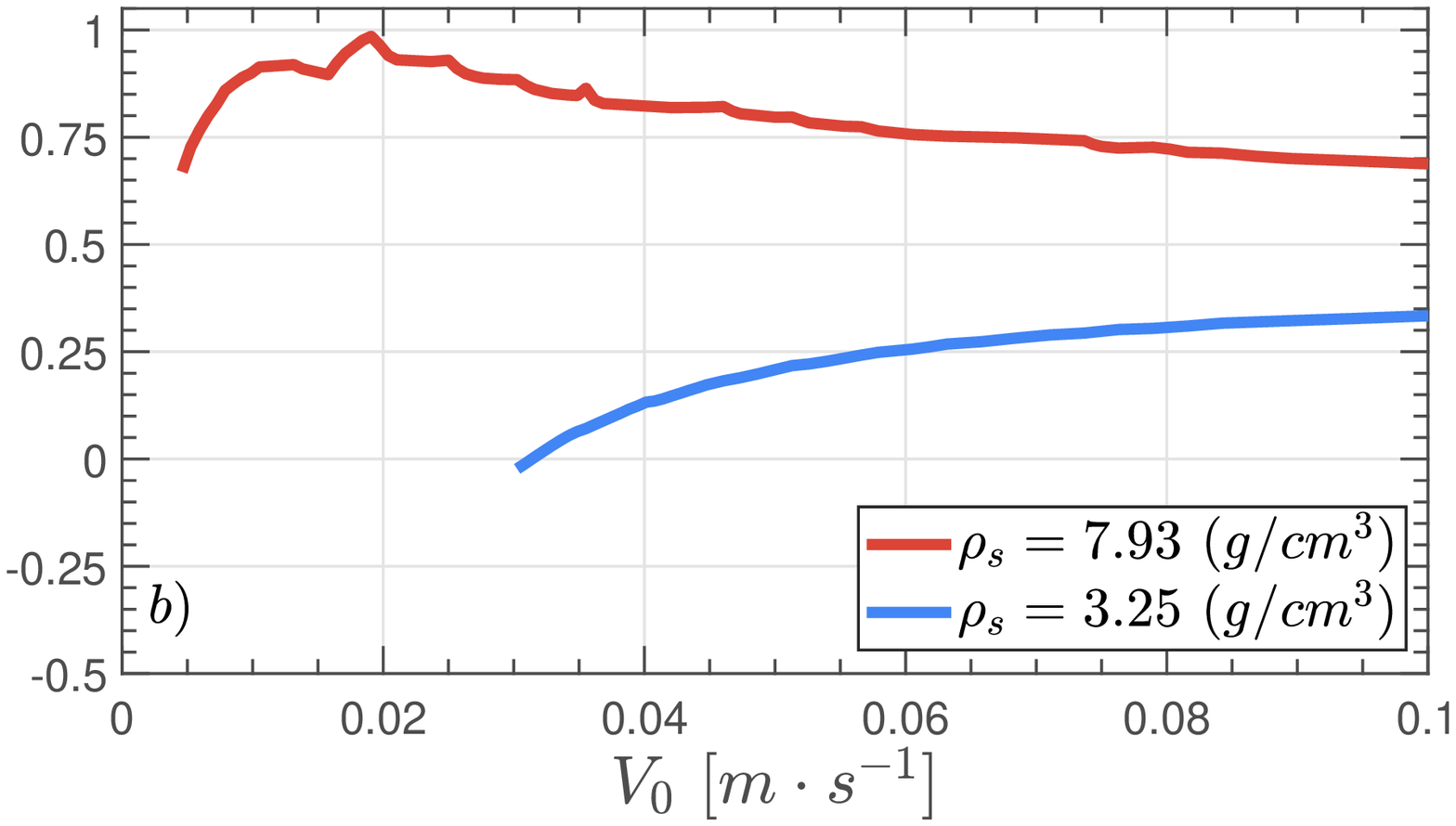}
    \end{tabular}
    \caption{Simulation results for the dependence of the coefficient of restitution squared on impact velocity for
    $\mathfrak{L} = 22.06$ and $\mathfrak{F} = 6.54 \times 10^{-5}$, with $\mathfrak{M} = 9.26 \times 10^{-3}$ (blue) and $\mathfrak{M} = 3.79 \times 10^{-3}$ (red). Impact velocities in panel (a) range from the minimum velocity required to produce a rebound to the maxima of the range explored by the experiments. Panel (b) zooms into the low $\mathfrak{U}$  limit.}
    \label{fig:Coeff_of_Restitution}
\end{figure}

We highlight that one should be careful to define the coefficient of restitution on a vertical bounce for which contact starts and ends at different heights (see the discussion in section 5.3 in \cite{Galeano-RiosEtAl2021}). If our interest is to quantify the transfer of energy during the rebound, the most adequate way to define the coefficient of restitution is
\begin{equation}
    \beta \coloneqq \sqrt{\frac{E^m_{\text{out}}}{E^m_{in}}},
\end{equation}
 where $E^m = E^k + E^p$, with $E^k$ being the kinetic energy of the sphere, and $E^p$ is the potential energy of the sphere measured with the zero reference level taken at $z = 1+\eta(r=0,t=0)$, i.e. the height of the centre of mass of the sphere when the impact starts. Moreover, the sub-indexes in $E^m_{\text{in}}$ and $E^m_{\text{out}}$ refer to the instants when landing and take-off are imminent, respectively. Notice, we define the end of contact differently for the purpose of this analysis.

It should be noted that, if the sphere never returns to the impact height (as is the case for the lowest impact velocity with $\rho_s = 3.25\,$g/cm$^3$), the definition above yields imaginary coefficients of restitution. To avoid imaginary numbers in the characterisation of our rebounds, we will instead use $\beta^2$ as our rebound metric, with the understanding that a negative value for $\beta^2$ implies a complete transfer of the initial energy of the sphere to the membrane plus an additional transfer to the membrane of the gravitational potential energy that the sphere had at the moment of first contact. 

We take $\mathfrak{L} = 22.06$ and $\mathfrak{F} = 6.54 \times 10^{-5}$, with $\mathfrak{M} = 9.26 \times 10^{-3}$ and $\mathfrak{M} = 3.79 \times 10^{-3}$ ; as in the experiments reported in figure \ref{fig:Comparison}, and we explore the low $\mathfrak{U}$ limit, going from the minimum velocities needed to produce a rebound to the maximum velocities used in the experiments. The dependence of $\beta^2$ on $\mathfrak{U}$ is reported in figure \ref{fig:Coeff_of_Restitution}. In \ref{fig:Coeff_of_Restitution}$a$, we can see that the coefficient of restitution depends weakly on the impact velocity only for the range of impact speeds explored in the experiments reported in figure \ref{fig:Comparison}. A different situation is observed in low-impact-velocity limit, amplified in figure \ref{fig:Coeff_of_Restitution}$b$, where there are clear changes in the fraction of energy that is recovered by the sphere, as the impact speed approaches the minimum speed for rebound. This behaviour is accompanied by a rise in the contact time as the minimum rebound speed is approached. This increase in contact time is by less than a factor of 2 in every case here considered. The behaviour of the system in this regime reflects the one observed for impacts on the free surface of a fluid bath for low Weber numbers (see figure 12b and 12e in \cite{Galeano-RiosEtAl2021}), thus establishing that this phenomenon is not exclusive to impacts occurring on a fluid.

We note that the curve that corresponds to $\rho_s = 7.93$ g/cm$^3$, in figure \ref{fig:Coeff_of_Restitution} shows some corners of numerical origin. These follow from the fact that, on a non-moving mesh, the contact area can only be approximated to the accuracy of the mesh, and therefore the radius of the contact area of some slightly different impacts will differ in their approximation by a full mesh interval. In cases of extremely weak impacts, the contact area can be so small that this error represents an appreciable fraction of the contact radius. That is, the jumps in the prediction reflect the jump in the approximations of the contact area. A similar effect was observed in the case of impacts on a fluid surface in \cite{Galeano-RiosEtAl2021} (see in figure 12.e the curve for $R = 0.25$mm). This effect is reduced with the used of a finer mesh; however, when using a uniform spatial mesh (as is the case here), this requires refining the mesh everywhere, which comes at an inconvenient computational cost. This problem will vanish with the introduction of a moving mesh, which is part of our ongoing work.

\subsubsection{Contact surface and pressure distribution}
\begin{figure}
    \centering
    \includegraphics[width=.7\textwidth]{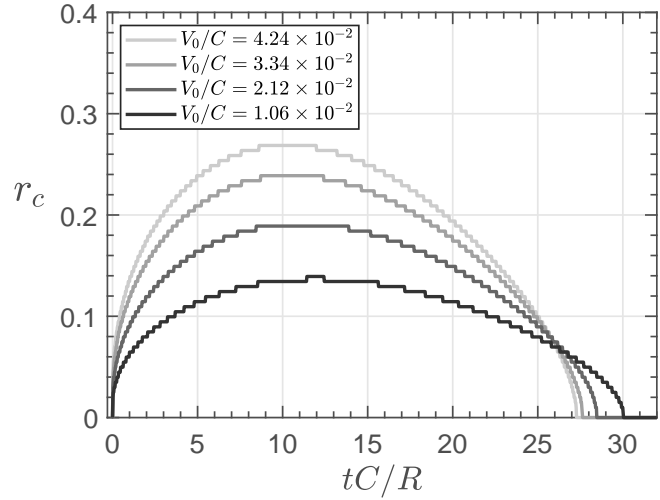}
    \caption{Simulation results for the evolution of dimensionless contact radius as a function of time for $\mathfrak{F}=6.54 \times 10^{-5}$, $\mathfrak{L}=22.06$, $\mathfrak{M} = 9.26 \times 10^{-3}$, for different values of $\mathfrak{U}= V_0/C$.} 
    \label{fig:pressed_area}
\end{figure}

The KM method also allows us to produce detailed predictions of the evolution of the contact surface and the pressure distribution on it. Figure \ref{fig:pressed_area} shows the evolution of the pressed radius as a function of time, for 
$\mathfrak{F}= 6.54 \times 10^{-5}$, $\mathfrak{L}=22.06$, $\mathfrak{M} = 9.26 \times 10^{-3}$; with four different experimental values of $\mathfrak{U} = V_0/C$.
The non-smooth nature of the curve shown in figure \ref{fig:pressed_area} is a natural consequence of our use of a non-moving mesh. A similar behaviour can be observed in the use of the KM for the case of impacts on fluids surfaces (see figures 8 and 9 in \cite{Galeano-RiosEtAl2017}).
\begin{figure}
    \centering
    \begin{tabular}{cc}
    \includegraphics[width=.45\textwidth]{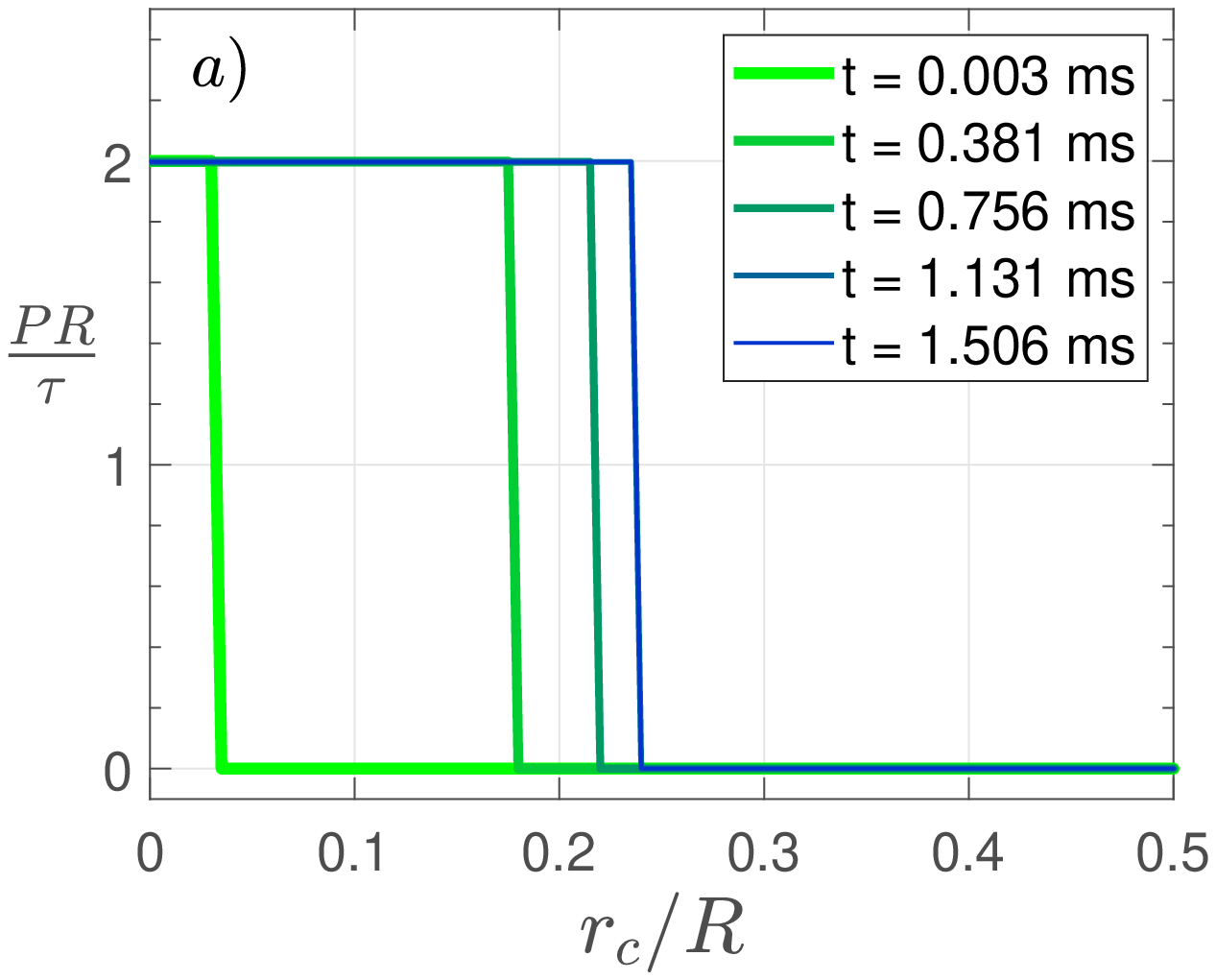}
         &  
    \includegraphics[width=.45\textwidth]{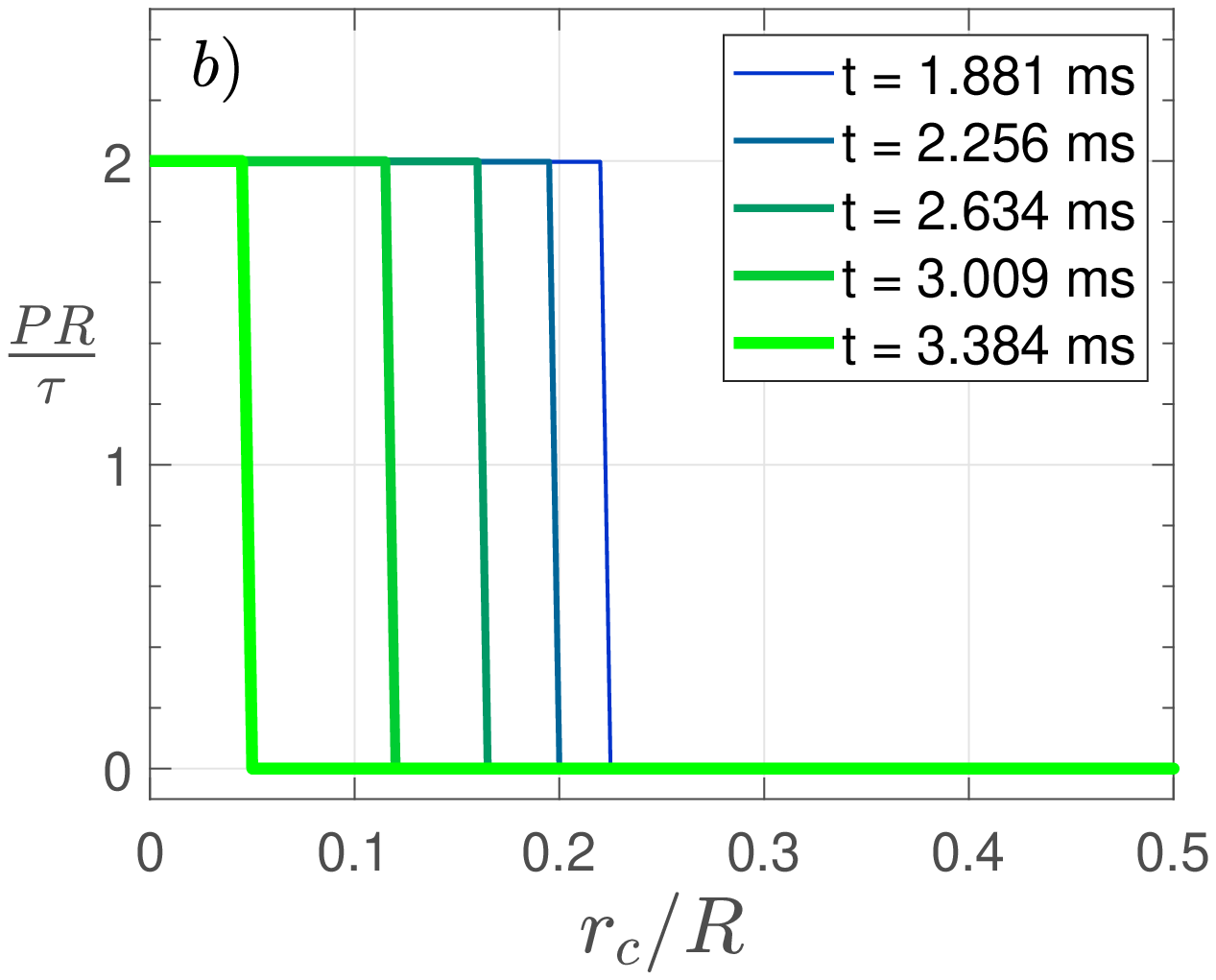}
    \end{tabular}
    \caption{Simulation results for pressure as a function of $r$ for different times during contact. Panel $a$ shows the evolution of the pressure field as the pressed area expands, and panel $b$ as it contracts. The case here presented corresponds to $\mathfrak{F}=6.54 \times 10^{-5}$, $\mathfrak{L}=22.06$, $\mathfrak{M} = 9.26 \times 10^{-3}$ and $\mathfrak{U}= 3.34 \times 10^{-2}$.}
    \label{fig:Pressure_distribution}
\end{figure}

The asymmetry in the curves in figure \ref{fig:pressed_area} is responsible for the transfer of momentum to the membrane. Indeed, if the membrane had no mass, the impact problem would become quasi-static, and the work done on the sphere by the membrane would be equal in magnitude, but of opposite sign, on the way down and the way up of the sphere, thus not allowing for energy transfer to the membrane.

Figure \ref{fig:Pressure_distribution} shows the evolution of the radial distribution of pressure for a typical rebound in these studies ($\mathfrak{F}=6.54 \times 10^{-5}$, $\mathfrak{L}=22.06$, $\mathfrak{M} = 9.26 \times 10^{-3}$ and $\mathfrak{U}= 3.34 \times 10^{-2}$). Panel $a$ shows the pressure distribution as the pressed area expands following first contact (panel $a$); and panel $b$ as it contracts on its way to take-off (panel $b$). It can be clearly seen that the pressure has an approximately constant value over the entire pressed area and throughout the duration of the impact. This value is, naturally, very close to the one given by the curvature
contribution (i.e. $2$ in the present non-dimensionalisation, see equations \ref{eqn:wave_non_lin_01} and \ref{eqn:check_kappa}); as, in the examples here considered, the inertia of the membrane beneath the sphere is very small, when compared to that of the sphere itself ($\mathfrak{M} \ll 1$). 

It is worth mentioning that, in the case of impacts on a fluid surface, the pressure underneath the impactor clearly shows a spike near the boundary of the pressed area as the impactor moves downward (see figure 8 in \cite{Galeano-RiosEtAl2021}, and figures 3g and 3f in \cite{HendrixEtAl2016}). The fact that here we do not observe such a spike in pressure is consistent with the claims made in \cite{Galeano-RiosEtAl2021}, which suggested such spikes are indeed caused by the fluid flow under the liquid surface.

\subsubsection{Multiplicity of contacts}
\begin{figure}
    \centering
    \includegraphics[width=.7\textwidth]{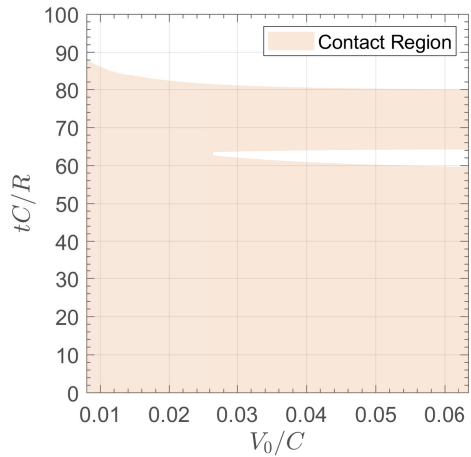}
    \caption{Evolution of the touch-down and take-off times as a function of initial velocity for $\mathfrak{F}= 1.81 \times 10^{-4}$, $\mathfrak{L}=16.54$, $\mathfrak{M} = 7.11 \times 10^{-3}$.} 
    \label{fig:mutiple_contacts}
\end{figure}

\begin{figure}
    \centering
    \includegraphics[width=\textwidth]{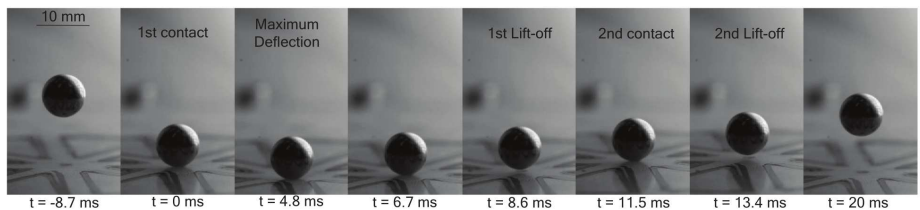}
    \caption{Double bounce as seen in experiments ($\mathfrak{L} = 13.08$, $\mathfrak{F} = 2.29 \times 10^{-4}$, $\mathfrak{M} = 5.62 \times 10^{-3}$ and $\mathfrak{U} = 8.23 \times 10^{-2}$).}
    \label{fig:double_contact_exp}
\end{figure}
While carrying out the investigations described above, we were also able to identify that, for certain parameter regimes, multiple contacts occur before the centre of the membrane moves downward a second time. Such double contacts were observed in simulations as well as in experiments. 

For $\mathfrak{F}= 1.81 \times 10^{-4}$, $\mathfrak{L}=16.54$, $\mathfrak{M} = 7.11 \times 10^{-3}$, we track the contact between the sphere and membrane in our simulations and we summarise the results in figure \ref{fig:mutiple_contacts}. The low $\mathfrak{U}$ limit does not show signs of multiple contacts. These appear at intermediate values of $\mathfrak{U} =V_0/C$, and the duration of the intermediate flight slowly increases with $\mathfrak{U}$.

We highlight that flights in between to contacts reported in figure \ref{fig:mutiple_contacts} are extremely short periods of mid-rebound flight, which are very difficult to measure in the experiments, and consequently it was not possible to verify these experimentally. Nevertheless, double contacts can be observed in some experiments for relatively higher $\mathfrak{U}$. Figure \ref{fig:double_contact_exp} shows one of these double contacts, observed in the experiments. A video of this double bounce is also made available as supplementary material, for a case in which this can be clearly seen. Unfortunately, these experimental rebounds with double contacts correspond to relatively strong impacts, which somewhat escape the linearity assumptions of our model, so a direct comparison was not realistic, and indeed our model did not predict a double rebound in the case for which it is was observed in the experiments.

We note that, when the sphere lifts off for the first time, the membrane enters a free oscillation regime; in which, the configuration of the membrane is described by a (potentially infinite) sum of standing modes, each with a different oscillation frequency. At the same time, the sphere is slowing down following lift-off, as fast oscillating modes in the membrane are to catch up with the south pole of the sphere once again.

\subsubsection{Non-monotonic decay of bouncing}
\begin{figure}
    \centering
    \includegraphics[width=\textwidth]{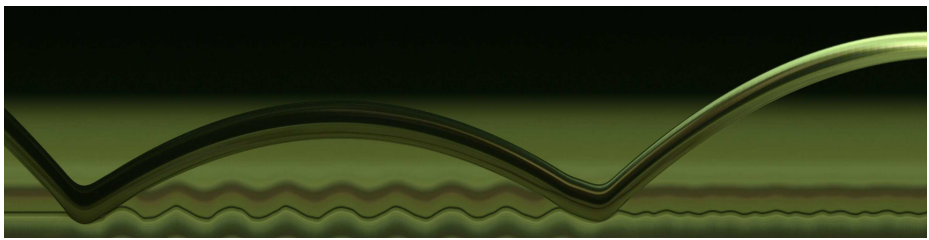}
    \caption{Spatiotemporal diagram composed of three-pixel-wide central slices for rebound experiment with $\mathfrak{L} = 13.08$, $\mathfrak{F} = 2.29 \times 10^{-4}$, $\mathfrak{M} = 5.62 \times 10^{-3}$ and $\mathfrak{U} = 8.39 \times 10^{-2}$.  Each slice is separated by 0.19 ms.  The sequence illustrates that a second impact can produce a coefficient of restitution greater than one, recovering previously transferred energy back from the vibrating membrane.}
    \label{fig:alpha_larger_than_1}
\end{figure}

\begin{figure}
    \centering
    \includegraphics[width=.8\textwidth]{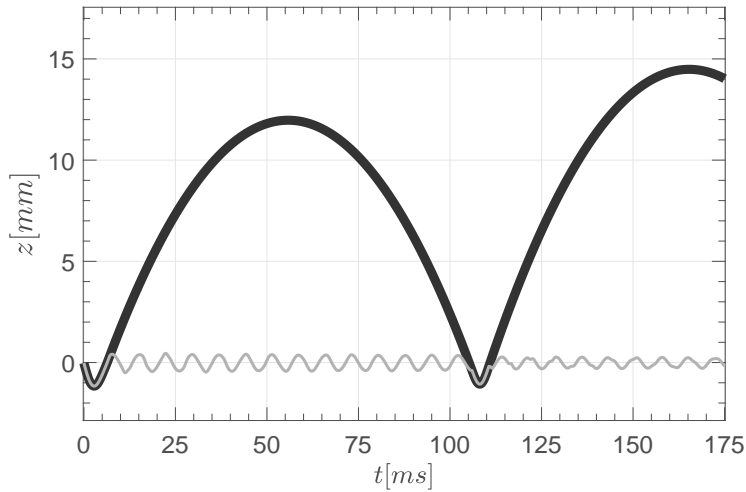}
    \caption{Example of second bounce with $\alpha > 1$ found in the simulations. The simulation corresponds to 
$\mathfrak{L} = 22.1987$, $\mathfrak{F} = 6.50 \times 10^{-5}$, $\mathfrak{M} = 3.82 \times 10^{-3}$ and $\mathfrak{U} = 3.62 \times 10^{-2}$.}
    \label{fig:alpha_larger_than_1_sim}
\end{figure}

Experimenting with somewhat stronger impacts, we are able to identify regimes in which a second rebound results in a coefficient of restitution that is greater than one ($\alpha>1$). In particular, this was observed in a case when the second impact happens as the centre of the membrane is moving downward (as if in phase with the impactor). Figure \ref{fig:alpha_larger_than_1}, which is constructed by placing three-pixel-wide central slices of the images on the bounce, illustrates this phenomenon. The figure clearly shows that during its second impact, the sphere is able to recover some of the energy it had bestowed to the membrane during the first bounce. A careful inspection of the first bounce in figure \ref{fig:alpha_larger_than_1} reveals that the phenomenon reported in figures \ref{fig:mutiple_contacts} and \ref{fig:double_contact_exp} is also present in this rebound. 

A second bounce with $\alpha>1$ can also be seen in the simulations (see figure \ref{fig:alpha_larger_than_1_sim}), proving that the model is able to capture this type of inertial effect. Moreover, we note that in all cases where we found $\alpha >1$, whether in experiments or in simulations, the sphere impacts the membrane as the centre of the membrane is moving downward, indicating that such an impact phase contributes to this effect. 

Figure \ref{fig:alpha_larger_than_1_sim} presents a sample case for which a second bounce with $\alpha>1$ is predicted in the simulations. A larger second bounce is also found in the experiments for these parameters. However, the second bounce is particularly sensitive to the first coefficient of restitution, as a differing flight time leads to a different impact phase. In particular, a direct quantitative comparison between model predictions and experimental results for this effect is currently impractical; as even a slight mismatch in the coefficient of restitution of the first impact (compounded by experimental uncertainty on the impact velocity) leads to different flight times (and impact phases) between consecutive bounces in the simulations. Regardless, this inertial effect is qualitatively observed in both experiment and simulation throughout wide parameter ranges. Even further experimental and modelling refinements would be necessary to achieve quantitative agreement, and will be the subject of future work.

\section{Discussion}\label{section:Discussion}

The application of the kinematic match method to the model problem here considered reveals the richness of what at first glance might appear as an exceedingly simple mechanical system. In particular, the possibility to model non-Hertzian effects allows us to capture behaviour that results from the wave-mediated exchanges of energy between the colliding solids.

The exploration of a parameter regime that lies away from the quasi-static limit (characterised by equations \ref{eqn:qs_scale} and \ref{eqn:qs_time_scales}) reveals new phenomena in both experiments and simulations. In particular, we have shown that the energy exchange with wave field in the membrane can lead to complex behaviour that includes multiple contacts over a single rebound, and non-monotonic decay of the rebound amplitude. These effects attest to the need for models of the present kind to have reasonable predictions for the coefficient of restitution, such as the ones given here.

For the range of impacts most readily accessible in experiment, the coefficient of restitution depends only weakly on impact velocity, it increases with the radius of the sphere, and it also increases for increased sphere density, when all other parameters are kept constant.

In the low impact velocity (low $\mathfrak{U}$), we observe the existence of a local maximum in the coefficient of restitution for certain radii, as well as negative values of the squared coefficient of restitution. To the best of our knowledge, these behaviours have never been identified in the impact of two solids. However, both effects have recently been shown to occur in impacts of solid spheres onto the free surface of a bath \cite{Galeano-RiosEtAl2021}. These findings are a strong indication that the analogous observations in impacts on fluids are not caused by fluid motion, but are instead a more general effect.

The predicted pressure distributions are approximately constant and equal to the curvature contribution over the pressed area. This is expected, as the mass of the membrane is small in the contact region in comparison to that of the sphere. The contrast between the pressure distribution found in this case and in the case of impacts on fluids, supports the claim made in \cite{Galeano-RiosEtAl2021}, that the peaks in the distribution observed in the case of impacts on fluids are caused by fluid motion effects.

We highlight also that the present work allowed us to improve the prior version of the kinematic match conditions. Moreover, this improvement will benefit the modelling of impacts on the free-surface of a fluid, as well all other future applications of the kinematic match.

In summary, the present work constitutes the first application of the kinematic match method for solids, while also providing experimental validation of results, and leading to the identification of previously unreported phenomena related to the transfer of energy in these impacts and multiple contacts over a single rebound event. Moreover, our work informs the study of analogous impacts on fluids by providing strong indications for causation of behaviours previously reported for those systems. Furthermore, the article improves the KM method, goes into extensive details on the technicalities involved in reproducing the calculations here presented, and it provides all code needed for independent verification and extension of our work.

\subsection{Future directions}
A closely related problem was studied by Eichwald and collaborators \cite{EichwaldEtAl2010}, who considered the case of a rigid sphere bouncing on a membrane that hermetically encloses a volume of air underneath it. The membrane and the air chamber below it are made to oscillate vertically, thus interacting with the bouncing sphere. They observed period doubling transitions into chaotic bouncing and related their results to the study of similar dynamical systems, in which a droplet bounces on the free-surface of a vibrating bath \cite{Couder2005_PRL} and for which the KM has already been used successfully \cite{Galeano-RiosEtAl2017,Galeano-RiosEtAl2019}. The exploration of the system presented in \cite{EichwaldEtAl2010}, using the fully predictive methods here introduced, is a potential natural extension of known applications of the KM.

Similar set-ups to those presented here and in \cite{CourbinEtAl2006} and \cite{EichwaldEtAl2010} were studied by Gilet and Bush in \cite{GiletAndBush2009,GiletAndBush2009PRL}, though in their case, the membrane was replaced by a fluid film and the sphere with a droplet. The methods here presented are also ideally suited to model system of this sort, provided the integrity of the fluid membrane is maintained during the rebound.

Our ongoing work includes the numerical implementation of the fully non-linear form of equation (\ref{eqn:wave_non_lin}), the use of higher order methods in time, the inclusion of the dependence of membrane tension on deformation (controlled by the elastic modulus), the determination of the threshold that separates impactors that rebound off the membrane after impact from those that no longer detach after the first contact, the detailed study of the coefficient of restitution, and an in-depth analysis of the rebounds that can recover energy from the membrane, as shown in figure \ref{fig:alpha_larger_than_1}.

Future ramifications of the present work also include the consideration of vibrating set-ups such as the ones covered in the work of \cite{EichwaldEtAl2010}, the study of impacts on fluid membranes, such as the one considered by \cite{GiletAndBush2009}, and the implementation of impacts by deformable spheres on deformable substrates. 

Furthermore, the problem of a rigid body impacting a deformable solid is clearly a free boundary problem. That is, the boundary $C(t)$ of the pressed surface $S(t)$ (see figure \ref{fig:Scheme}) is an unknown curve that separates two regions where different partial differential equations are to be solved. Therefore, it is more appropriate to treat the problem with a moving mesh whose nodes are able to follow the moving boundary exactly. A finite element method implementation that uses the spine method to track moving boundaries is being developed to more adequately impose the matching conditions in system (\ref{eqn:eta_t})-(\ref{eqn:no_overlap}). This implementation is of particular interest, given that it can be generalised to manage non-symmetric impacts in a natural way. Moreover, with only a few changes, the method can be adapted to model impacts in which both impacting bodies deform, greatly increasing the number of potential engineering applications of it.

Given that the KM is agnostic in relation to the form of the equations that govern the behaviour of the impacting surfaces, the methods here implemented have great potential for broader applications. In particular, this opens the possibility to use it to model inelastic collisions, and the plastic deformation that comes with them, in a fully predictive way and in very general set-ups.

\appendix

\section{Governing equation for the elastic membrane}\label{App:Wave_eq}

\begin{figure}
    \centering
    \includegraphics[width = .4\textwidth]{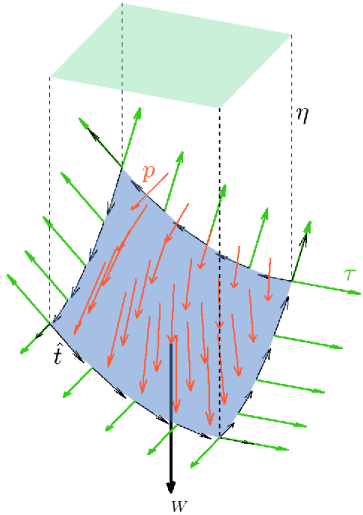}
    \caption{Schematic representation of the forces acting on an arbitrary element of the elastic membrane for a case with negative surface elevation ($\eta<0$) and membrane-element weight $W = \mu g \mathscr{A}$.}
    \label{fig:Scheme_derivation}
\end{figure}

For simplicity, we derive the equations in Cartesian coordinates. We recall our assumption that the membrane deforms exclusively on the $z$ direction, and thus we consider the dimensional vertical deflection of the membrane $\eta (x,y,t)$ in an arbitrary surface element $\mathscr{S}$, whose projection on the $(x,y)$-plane is given by $\mathscr{A}$. Thus, the $z$-component of Newton's second law of motion for this arbitrary membrane element in dimensional form is given by 
\begin{equation}\label{eqn:apx_newton1}
    \mu
    \int\limits_{\mathscr{A}}
    \partial_{tt} \eta \  d\mathscr{A} 
    =
    -g\mu \int\limits_{\mathscr{A}}\   d\mathscr{A} 
    -
    \int\limits_{\mathscr{S}} p \inner{\hat{n}}{\hat{z}}  d\mathscr{S} 
    + 
    \tau 
    \int\limits_{\mathscr{C}} \inner{\hat{t} \times \hat{n}}{\hat{z}} dl,
\end{equation}
where $\mathscr{C}$ is the boundary of $\mathscr{S}$, $\hat{z}$ is the unit vector pointing in the direction of the $z$ axis, and we recall that $\mu$ is the mass per unit area of the membrane, which is assumed to be constant, $p = p(x, y)$ is the pressure distribution on top of the membrane (which is
positive when it points into the membrane), $\tau$ is the isotropic stress of the membrane (i.e. normal force per unit length), which is also assumed constant in time and space, $\hat{n}$ is the upward-pointing unit vector that is normal to the membrane, and $\hat{t}$ is the unit tangent vector to the curve $\mathscr{C}$, which is oriented so that $\hat{t}\times\hat{n}$ points in the direction of the membrane traction stress (see figure \ref{fig:Scheme_derivation}).

The left-hand side of (\ref{eqn:apx_newton1}) corresponds to the mass of the surface element times the acceleration of the centre of mass while the right-hand side has the contribution of the downward force exerted by gravity, the normal forces due to the pressure distribution, and the isotropic stress (which is pulling away from the membrane element in the direction that is tangent to the membrane and normal to $\mathscr{C}$), respectively. 

We define $F(x,y,z) = z-\eta(x,y,t)$; and, consequently, 
\begin{equation}
    \hat{n}(x,y) = \dfrac{\nabla F}{|\nabla F|},
\end{equation}
i.e.
\begin{equation}\label{eqn:apx_normal}
    \hat{n}
    =
    \frac{
    \left(-\partial_x \eta,
    -\partial_y \eta, 1\right)^T
    } % End numerator
    {
    \sqrt{\left(\partial_x \eta \ \right)^2
    +
    \left(\partial_y \eta \ \right)^2
    +1}
    }. % End denominator
\end{equation}
Moreover
\begin{equation} 
\label{eqn:apx_pressure}
    \int\limits_{\mathscr{S}} p \inner{\vec{n}}{\vec{z}}  d\mathscr{S}
    =
    \int\limits_{\mathscr{A}} p \ d\mathscr{A},
\end{equation}
and we have
\begin{equation}
    \inner{\hat{t} \times \hat{n}}{\hat{z}}    
    =
    \frac{
    \inner{\hat{t}}{\hat{y}}
    \partial_x \eta 
    - 
    \inner{\hat{t}}{\hat{x}}
    \partial_y \eta
    }
    {
    \sqrt{\left(\partial_x \eta\ \right)^2
    +
    \left(\partial_y \eta\ \right)^2 + 1}
    },
\end{equation}
where $\hat{x}$ and $\hat{y}$ are the unit vectors that point in the direction of the $x$ and $y$ axes.

Defining an auxiliary vector field $\vec{w}$, so that $\vec{w}: (x, y, z) \mapsto  \left(-\partial_y \eta,  \partial_x \eta, 0\right) / |
\nabla F|$, which is smooth in a neighbourhood of $\mathscr{S} \subset \mathbb{R}^{3}$, we can express the line integral in \ref{eqn:apx_newton1} as 
\begin{equation} \label{eqn:apx_stokes}
    \int\limits_{\mathscr{C}} \left< \hat{t} \times \hat{n}, \hat{z} \right> dl
    =
    \int\limits_{\mathscr{C}} \vec{w} \cdot \vec{dl} 
    = 
    \int\limits_{\mathscr{S}}\inner{\nabla \times \vec{w}}{\hat{n}} d\mathscr{S},
\end{equation}
where the last equality follows from Stokes' theorem. 

We note that
\begin{equation}
    \nabla \times \vec{w} = \left(0,0,\partial_x\frac{\partial_x\eta}{\sqrt{\left(\partial_x\eta\right)^2+\left(\partial_y\eta\right)^2+1}}+\partial_y\frac{\partial_x\eta}{\sqrt{\left(\partial_x\eta\right)^2+\left(\partial_y\eta\right)^2+1}}\right)^T,
\end{equation}
and, therefore, we have
\begin{equation}
    \inner{\nabla \times \vec{w}}{\hat{n}}
    =
    \left(\nabla \cdot \hat{n}\right) \inner{\hat{n}}{\hat{z}},
\end{equation}
which implies
\begin{equation}
    \label{eqn:apx_divergence}
    \int\limits_{\mathscr{S}}
    \inner{\nabla \times \vec{w}}{\hat{n}} d\mathscr{S}
    =
    \int\limits_{\mathscr{S}} 
    \left(\nabla \cdot \hat{n}\right)
    \inner{\hat{n}}{\hat{z}}
    \ d\mathscr{S}
    = 
    \int\limits_{\mathscr{A}} 
    \left(\nabla \cdot \hat{n}\right)
    \ d\mathscr{A},
\end{equation}
where $\kappa(\eta) = \nabla\cdot\hat{n}$ is exactly twice the mean curvature operator.

Together, equations (\ref{eqn:apx_newton1}), (\ref{eqn:apx_pressure}), (\ref{eqn:apx_divergence}) allow us to write

\begin{equation} \label{eqn:apx_integralform}
    \int\limits_{\mathscr{A}} 
    \mu \partial_{tt} \eta  \ d\mathscr{A} 
    =
    \int\limits_{\mathscr{A}}
    \left( -g \mu
    +
    \tau \kappa (\eta)
    -
    p
    \right)
    \ d\mathscr{A}.
\end{equation}

Now, since $\mathscr{A}$ is arbitrary, we must have
\begin{equation}
\label{eqn:governing_membrane}
    \mu \partial_{tt} \eta 
    =
    -g \mu
    +
    \tau \kappa (\eta)
    -
    p.
\end{equation}

Finally, we note that the steady state version of equation (\ref{eqn:governing_membrane}), used for the initial condition is obtained when the left hand side and the pressure term are both equal to 0.

\section{Algorithm}\label{sec:Algorithm}
A general pseudocode used as a template for the different implementations of our method is given in algorithm \ref{alg:IterationOnGeometry}.

\begin{algorithm}[H] \label{alg:IterationOnGeometry}
    \DontPrintSemicolon
    \SetKwInOut{Input}{Input}
    \SetKwInOut{Output}{Output}
    \Begin{
     InitialiseProblemConditions();\;
    int i = 0; int j = 0; \;
    \While{simulationTime $\leq$ finalTime}{
     recalculate \hspace{2pt} $\longleftarrow False$;\;
     oneLess \hspace{12pt} $\longleftarrow$ 
     tryAdvancingOneStep($\delta_{t}^{k+1}$, contactPoints $- 1$);\;
     samePoints $\longleftarrow$ 
     tryAdvancingOneStep($\delta_{t}^{k+1}$, contactPoints);\;
     oneMore \hspace{8pt} $\longleftarrow$ 
     tryAdvancingOneStep($\delta_{t}^{k+1}$, contactPoints $+1$);\;
     minimum \hspace{4pt} $\longleftarrow$ 
     min(oneLess.error, samePoints.error, oneMore.error);\;
     
     \uIf{minimum $==$ samePoints.error}{
        currentVariables $\longleftarrow$ samePoints;\;
     } 
     \uElseIf{minimum $==$ oneMore.error}{
     twoMore $\longleftarrow$  tryAdvancingOneStep($\delta_{t}^{k+1}$, contactPoints $+ 2$);\;
         \eIf{oneMore.error $<$ twoMore.error}
            {currentVariables $\longleftarrow$ oneMore;\;}
            {recalculate $\longleftarrow True$;\;}
     } \uElseIf{minimum $==$ oneLess.error}{
     twoLess $\longleftarrow$  tryAdvancingOneStep($\delta_{t}^{k+1}$, contactPoints $- 2$);\;
         \eIf{oneLess.error $<$ twoLess.error}
            {currentVariables $\longleftarrow$ oneLess;\;}
            {recalculate $\longleftarrow True$;\;}
     }
     \eIf{recalculate $ == True$}
        {$\delta_{t}^{k+1} \longleftarrow \delta_{t}^{k+1}/2$;
        $\ i \longleftarrow i + 1$; $\ j \longleftarrow 2 j$ ;}
        {
        $simulationTime \longleftarrow simulationTime + \delta_{t}^{k+1}$;
        $\ j \longleftarrow j + 1$;\;
        \If{$j\%2 == 0$ }
            {$\delta_{t}^{k+1} \longleftarrow 2 \delta_{t}^{k+1}$;  
            $\ i \longleftarrow i - 1$;
            $\ j \longleftarrow j/2$;}
        }
        \If{$2^i == j$}
        { $j \longleftarrow 0$;}
    } % End while
    } % End Begin
    \caption{Pseudocode for the algorithm implemented in this work}
\end{algorithm}
\vskip6pt

\enlargethispage{20pt}

% \ethics{No ethical concerns to declare.}
Data Access: {All simulation parameters are given in the text and all simulation codes necessary to reproduce the results are provided as through a public repository.}

Author contribution: {EAAV and CAGR developed the mathematical model and performed the simulations. LA and DMH designed and performed the experiments. All authors contributed to the writing of the manuscript.}\newline
{The authors declare no competing interests.}\newline
{CAGR gratefully acknowledges the support of EPSRC project EP/P031684/1.}

% \ack{Insert acknowledgement text here.}

% \disclaimer{Insert disclaimer text here.}
%%%%%%%%%% Insert bibliography here %%%%%%%%%%%%%%
\bibliographystyle{RS.bst}
\bibliography{Biblio_Impacts.bib}

\begin{thebibliography}{99}

\bibitem{bruzzone20212d}
Bruzzone F, Maggi T, Marcellini C, Rosso C. 2021a  2D Nonlinear and
  non-Hertzian gear teeth deflection model for static transmission error
  calculation. {\em Mechanism and Machine Theory} \textbf{166}, 104471.

\bibitem{bruzzone2021gear}
Bruzzone F, Maggi T, Marcellini C, Rosso C. 2021b  Gear Teeth Deflection Model
  for Spur Gears: Proposal of a 3D Nonlinear and Non-Hertzian Approach. {\em
  Machines} \textbf{9}, 223.

\bibitem{rosso2019proposal}
Rosso C, Bruzzone F, Maggi T, Marcellini C. 2019  A proposal for
  semi-analytical model of teeth contact with application to gear dynamics.
  Technical report SAE Technical Paper.

\bibitem{askari2021mathematical}
Askari E. 2021  Mathematical models for characterizing non-Hertzian contacts.
  {\em Applied Mathematical Modelling} \textbf{90}, 432--447.

\bibitem{hundal1976impact}
Hundal M. 1976  Impact absorber with linear spring and quadratic law damper.
  {\em Journal of Sound and Vibration} \textbf{48}, 189--193.

\bibitem{Johnson1987}
Johnson KL, Johnson KL. 1987 {\em Contact mechanics}.
Cambridge university press.

\bibitem{cundall1979discrete}
Cundall PA, Strack OD. 1979  A discrete numerical model for granular
  assemblies. {\em geotechnique} \textbf{29}, 47--65.

\bibitem{herbert2001measurement}
Herbert E, Pharr G, Oliver W, Lucas B, Hay J. 2001  On the measurement of
  stress--strain curves by spherical indentation. {\em Thin solid films}
  \textbf{398}, 331--335.

\bibitem{begley2004spherical}
Begley MR, Mackin TJ. 2004  Spherical indentation of freestanding circular thin
  films in the membrane regime. {\em Journal of the Mechanics and Physics of
  Solids} \textbf{52}, 2005--2023.

\bibitem{komaragiri2005mechanical}
Komaragiri U, Begley M, Simmonds J. 2005  The mechanical response of
  freestanding circular elastic films under point and pressure loads. {\em J.
  Appl. Mech.} \textbf{72}, 203--212.

\bibitem{selvadurai2006deflections}
Selvadurai A. 2006  Deflections of a rubber membrane. {\em Journal of the
  Mechanics and Physics of Solids} \textbf{54}, 1093--1119.

\bibitem{ahearne2005characterizing}
Ahearne M, Yang Y, El~Haj AJ, Then KY, Liu KK. 2005  Characterizing the
  viscoelastic properties of thin hydrogel-based constructs for tissue
  engineering applications. {\em Journal of the Royal Society Interface}
  \textbf{2}, 455--463.

\bibitem{gupta2015recent}
Gupta A, Sakthivel T, Seal S. 2015  Recent development in 2D materials beyond
  graphene. {\em Progress in Materials Science} \textbf{73}, 44--126.

\bibitem{Hertz1896}
Hertz H. 1896 {\em Miscellaneous papers,(ed.) P. Lenard}.
Macmillan London.

\bibitem{Hertz1882beruhrung}
Hertz H. 1882  {\"U}ber die Ber{\"u}hrung fester elastischer K{\"o}rper. {\em
  Journal f{\"u}r die reine und angewandte Mathematik} \textbf{92}, 22.

\bibitem{Johnson1982}
Johnson KL. 1982  One hundred years of Hertz contact. {\em Proceedings of the
  Institution of Mechanical Engineers} \textbf{196}, 363--378.

\bibitem{Karami1989non}
Karami G. 1989  Non-Hertzian Contact Problems. In {\em Lecture Notes in
  Engineering} pp. 145--203. Springer.

\bibitem{KalkerAndRanden1972}
Kalker J, Van~Randen Y. 1972  A minimum principle for frictionless elastic
  contact with application to non-Hertzian half-space contact problems. {\em
  Journal of engineering mathematics} \textbf{6}, 193--206.

\bibitem{PaulAndHashemi1981}
Paul B, Hashemi J. 1981  {Contact Pressures on Closely Conforming Elastic
  Bodies}. {\em Journal of Applied Mechanics} \textbf{48}, 543--548.

\bibitem{LeeAndKim2008}
Lee DG, Kim HY. 2008  Impact of a superhydrophobic sphere onto water. {\em
  Langmuir} \textbf{24}, 142--145.

\bibitem{CourbinEtAl2006}
Courbin L, Marchand A, Vaziri A, Ajdari A, Stone HA. 2006  Impact dynamics for
  elastic membranes. {\em Physical review letters} \textbf{97}, 244301.

\bibitem{EichwaldEtAl2010}
Eichwald B, Argentina M, Noblin X, Celestini F. 2010  Dynamics of a ball
  bouncing on a vibrated elastic membrane. {\em Physical Review E} \textbf{82},
  016203.

\bibitem{Galeano-RiosEtAl2017}
Galeano-Rios CA, Milewski PA, Vanden-Broeck JM. 2017  Non-wetting impact of a
  sphere onto a bath and its application to bouncing droplets. {\em Journal of
  Fluid Mechanics} \textbf{826}, 97--127.

\bibitem{Galeano-RiosEtAl2019}
Galeano-Rios CA, Milewski PA, Vanden-Broeck JM. 2019  Quasi-normal free-surface
  impacts, capillary rebounds and application to Faraday walkers. {\em Journal
  of Fluid Mechanics} \textbf{873}, 856--888.

\bibitem{Galeano-RiosEtAl2021}
Galeano-Rios CA, Cimpeanu R, Bauman IA, MacEwen A, Milewski PA, Harris DM. 2021
   Capillary-scale solid rebounds: experiments, modelling and simulations. {\em
  Journal of Fluid Mechanics} \textbf{912}.

\bibitem{GiletAndBush2009}
Gilet T, Bush JW. 2009  The fluid trampoline: droplets bouncing on a soap film.
  {\em Journal of Fluid Mechanics} \textbf{625}, 167--203.

\bibitem{nagurka2004mass}
Nagurka M, Huang S. 2004  A mass-spring-damper model of a bouncing ball. In
  {\em Proceedings of the 2004 American control conference} vol.~1 pp.
  499--504. IEEE.

\bibitem{HendrixEtAl2016}
Hendrix MH, Bouwhuis W, van~der Meer D, Lohse D, Snoeijer JH. 2016  Universal
  mechanism for air entrainment during liquid impact. {\em Journal of fluid
  mechanics} \textbf{789}, 708--725.

\bibitem{Couder2005_PRL}
Couder Y, Fort E, Gautier CH, Boudaoud A. 2005  From bouncing to floating:
  noncoalescence of drops on a fluid bath. {\em Physical review letters}
  \textbf{94}, 177801.

\bibitem{GiletAndBush2009PRL}
Gilet T, Bush JW. 2009  Chaotic bouncing of a droplet on a soap film. {\em
  Physical review letters} \textbf{102}, 014501.

\end{thebibliography}

\end{document}